\documentclass[10pt,twocolumn,showpacs,amsmath,amssymb,floatfix,superscriptaddress]{revtex4-2}
\usepackage{graphicx}
\usepackage{float}
\usepackage{dcolumn}
\usepackage{bm}
\usepackage{color}
\usepackage{txfonts}
\usepackage{microtype}
\usepackage{hyperref}
\usepackage{mathrsfs}
\usepackage{easyReview}
\usepackage{amsmath}
\usepackage{extarrows}
\usepackage{mathtools}
\usepackage{xcolor}
\usepackage{slashed}
\usepackage{amsmath}

\setreviewson 
\begin{document}
\title{Manipulation of Superposed Vortex States of $\gamma$ Photon via Nonlinear Compton Scattering}
    
	\author{Jun-Lin Zhou}
	\affiliation{Ministry of Education Key Laboratory for Nonequilibrium Synthesis and Modulation of Condensed Matter, State key laboratory of electrical insulation and power equipment, Shaanxi Province Key Laboratory
    of Quantum Information and Quantum Optoelectronic Devices, School of Physics, Xi’an Jiaotong University, Xi’an 710049, China}
	\author{Mamutjan Ababekri}\email{mamutjan@xjtu.edu.cn}
	\affiliation{Ministry of Education Key Laboratory for Nonequilibrium Synthesis and Modulation of Condensed Matter, State key laboratory of electrical insulation and power equipment, Shaanxi Province Key Laboratory
    of Quantum Information and Quantum Optoelectronic Devices, School of Physics, Xi’an Jiaotong University, Xi’an 710049, China}
    \author{Yong-Zheng Ren}
	\affiliation{Ministry of Education Key Laboratory for Nonequilibrium Synthesis and Modulation of Condensed Matter, State key laboratory of electrical insulation and power equipment, Shaanxi Province Key Laboratory
    of Quantum Information and Quantum Optoelectronic Devices, School of Physics, Xi’an Jiaotong University, Xi’an 710049, China}
	\author{Yu Wang}
	\affiliation{Ministry of Education Key Laboratory for Nonequilibrium Synthesis and Modulation of Condensed Matter, State key laboratory of electrical insulation and power equipment, Shaanxi Province Key Laboratory
    of Quantum Information and Quantum Optoelectronic Devices, School of Physics, Xi’an Jiaotong University, Xi’an 710049, China}
    \author{Ren-Tong Guo}
	\affiliation{Ministry of Education Key Laboratory for Nonequilibrium Synthesis and Modulation of Condensed Matter, State key laboratory of electrical insulation and power equipment, Shaanxi Province Key Laboratory
    of Quantum Information and Quantum Optoelectronic Devices, School of Physics, Xi’an Jiaotong University, Xi’an 710049, China} 
    \author{Zhao-Hui Chen}
	\affiliation{Ministry of Education Key Laboratory for Nonequilibrium Synthesis and Modulation of Condensed Matter, State key laboratory of electrical insulation and power equipment, Shaanxi Province Key Laboratory
    of Quantum Information and Quantum Optoelectronic Devices, School of Physics, Xi’an Jiaotong University, Xi’an 710049, China}
	\author{Yu-Han Kou}
	\affiliation{Ministry of Education Key Laboratory for Nonequilibrium Synthesis and Modulation of Condensed Matter, State key laboratory of electrical insulation and power equipment, Shaanxi Province Key Laboratory
    of Quantum Information and Quantum Optoelectronic Devices, School of Physics, Xi’an Jiaotong University, Xi’an 710049, China}
    \author{Zhong-Peng Li}
	\affiliation{Ministry of Education Key Laboratory for Nonequilibrium Synthesis and Modulation of Condensed Matter, State key laboratory of electrical insulation and power equipment, Shaanxi Province Key Laboratory
    of Quantum Information and Quantum Optoelectronic Devices, School of Physics, Xi’an Jiaotong University, Xi’an 710049, China}
	\author{Jian-Xing Li}\email{jianxing@xjtu.edu.cn}
	\affiliation{Ministry of Education Key Laboratory for Nonequilibrium Synthesis and Modulation of Condensed Matter, State key laboratory of electrical insulation and power equipment, Shaanxi Province Key Laboratory
    of Quantum Information and Quantum Optoelectronic Devices, School of Physics, Xi’an Jiaotong University, Xi’an 710049, China}
	\affiliation{Department of Nuclear Physics, China Institute of Atomic Energy, P. O. Box 275(7), Beijing 102413, China}
	\date{\today}

\begin{abstract}
Vortex $\gamma$ photons in superposition states have important applications in photonuclear, high-energy, and strong-field physics. However, their controlled generation in the $\gamma$-ray regime remains a great challenge. Here, we put forward a novel method for the generation of vortex $\gamma$ photon in superposition states, with controllable orbital angular momentum (OAM) separation $\Delta\ell^\prime$ and modal weights, via nonlinear Compton scattering driven by multifrequency circularly polarized laser fields. We develop a strong-field quantum electrodynamics (QED) framework to reveal the underlying mechanism and calculate the radiation probabilities. In our method, the superposition arises from interference between energy-degenerate multiphoton pathways carrying distinct OAM. For two-frequency fields, the OAM separation follows $\Delta\ell'=\nu\mp1$ (upper/lower sign for equal/opposite helicities), and modal weights are tunable by laser intensities, with $\nu$ the frequency ratio. 
Vortex $\gamma$ photons in controllable superposition states from our method have significant applications in strong-field QED and nuclear photonics.
\end{abstract}
  
\maketitle
Controllable generation of vortex photons in orbital angular momentum (OAM) superposition states has broad applications in structured-light research; compared with single-OAM eigenstates, these states offer higher-dimensional information capacity and enhanced quantum control \cite{Allen:1992zz,Shen2019,Forbes:2021tpp,forbes2025progress}. In such states, multiple OAM components coexist for the same photon energy and helicity, making their relative phase directly accessible via interference \cite{mairEntanglementOrbitalAngular2001,Vaziri_2002}. By exploiting high-dimensional Hilbert spaces \cite{MolinaTerriza2007,Yao2011,erhard2020advances}, these states enable enhanced quantum control, increased information capacity, and modified light--matter selection rules \cite{kapaleVortexPhaseQubit2005,Moxley:2015yil,Babazadeh:2017ioi,Huang:2025jqa}. In the $\gamma$-ray domain, coherent OAM superposition causes nuclear level interference, altering transitions and excitation probabilities. It generates azimuthal absorption cross-section distributions \cite{Lu:2023wrf,Ujeniuc:2024hse,Xu:2024jlt} and other phenomena, paving new ways for OAM spectroscopy, nuclear structure \cite{Klein2019,Ivanov:2022jzh,Zilges:2022ugq,5md2-ngcf2026}, novel quantum electrodynamics (QED) effects \cite{PhysRevA.96.062120,PhysRevLett.123.113604,PhysRevResearch.3.043159} and high-energy OAM detection \cite{Ivanov:2019vxe,afanasev2021recoil}.
Vortex states with fixed $\Delta\ell'$ produce distinct $\Delta\ell'$-fold interference fringes \cite{PhysRevA.91.013403,xiaoGenerationPhotonicOrbital2016,knyazev2018beams}. As intrinsic signatures of photon spatial structure, they are observable even with unresolved absolute OAM, overcoming identification limitations in conventional scattering \cite{PhysRevA.92.013401,Ababekri:2024cyd}. Therefore, the central challenge is how to controllably prepare vortex $\gamma$ photon in coherent superpositions of distinct OAM modes.

Research on superposed vortex photons has advanced across many lower‑energy platforms, including optical beam shaping, metasurfaces, integrated photonics, matter waves, soft x‑rays, and free‑electron lasers  \cite{vasilyeu_generating_2009,liPatternManipulationOnchip2015,RubinszteinDunlop2017,lee_laguerregauss_2019,yao_alldielectric_2021,liu_generation_2024}. 
In these regimes, superposition states are typically prepared via external spatial‑mode shaping, splitting, or recombination.  
In the $\gamma$-ray regime, existing work has largely produced OAM-carrying $\gamma$ photons \cite{Jentschura:2010ap,petrilloComptonScatteredXGamma2016,taira2017gamma,Ababekri:2022mob,PhysRevLett.134.153802}, but controllable preparation of vortex $\gamma$ photon in superposition states at fixed kinematics remains unrealized. Because modes cannot be reshaped after emission, such superpositions must be encoded and tunably controlled at the emission vertex, making their preparation a major challenge.

Existing approaches to structured $\gamma$ photon generation range from plasma-based mechanisms to strong-field laser--electron interactions \cite{Chen:2018tkb,liuVortexRaysScattering2020,wangGenerationIntenseVortex2020,huAttosecondRayVortex2021,zhangEfficientBrightGray2021,bakeBrightGraySource2022,Bu2024,Wei:2025zsv,Liual._2026}. Nonlinear Compton scattering (NCS) and related processes can generate vortex photons carrying intrinsic OAM through angular-momentum conservation  \cite{bahrdt2013first,katoh2017angular,Bogdanov:2019ocq,chen2019generation,maruyama2025photon,liao2025all}. 
In single-color pulsed circularly polarized (CP) fields, the finite spectral bandwidth leads to slight overlap between adjacent harmonics, which is typically restricted to $\Delta\ell'=1$ \cite{Ababekri:2022mob}, and thus cannot provide a general control principle for superposition states synthesis.
This naturally motivates multifrequency drivers, where channel degeneracy can be engineered rather than merely encountered \cite{narozhny2000quantum,PhysRevD.90.125008,PhysRevA.98.052130}. Recent multicolor schemes have demonstrated control over polarization and vortex charge \cite{PhysRevLett.134.153802}. Meanwhile, semiclassical studies of multifrequency undulators have shown that engineered electron trajectories can emit photons in composite twisted states \cite{Bogdanov:2025vdl,Bogdanov:2026tbl}. 
However, the controllable generation of vortex $\gamma$ photon in superposition states at the radiation vertex via strong-field QED mechanisms in the high-energy regime remains a critical challenge.

\begin{figure*}
		\setlength{\abovecaptionskip}{-0.00cm}
		\includegraphics[width=0.975\linewidth]{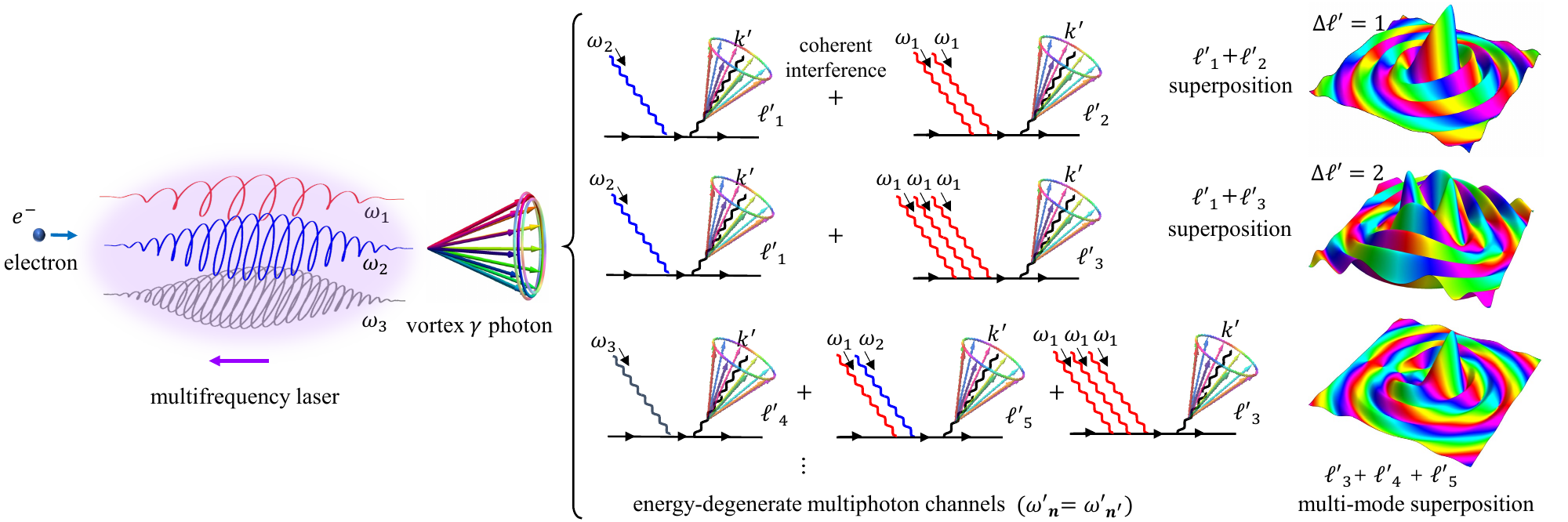}
		\begin{picture}(300,0)	
		\end{picture}
        \caption{Schematic illustration of the generation of vortex $\gamma$ photon in superposition states via NCS driven by the multifrequency laser. Left: A relativistic electron collides with a counterpropagating multifrequency CP laser field and emits a $\gamma$ photon. Middle: Energy-degenerate multiphoton channels with different angular momentum interfere at the emission vertex, producing coherent vortex superpositions. Examples for $\Delta\ell'=1$ (top), $\Delta\ell'=2$ (middle), and higher-order channel combinations (bottom).  Right: Representative transverse intensity and phase profiles of the resulting Bessel OAM superposition states.}
		\label{fig_1}
\end{figure*}
\begin{figure}
		\setlength{\abovecaptionskip}{-0.0cm}
		\includegraphics[width=0.995\linewidth]{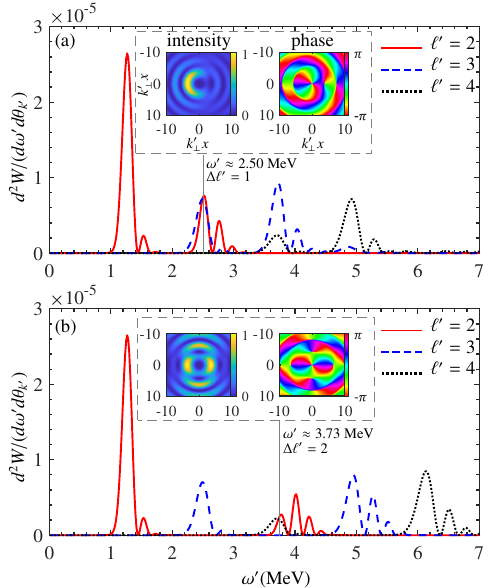}
		\begin{picture}(300,0)
		\end{picture}
\setlength{\abovecaptionskip}{-0.5 cm}
\caption{Energy-resolved emission probability of vortex $\gamma$ photons in NCS driven by two-frequency lasers. The emitted photon helicity is $\Lambda'=-1$, with laser intensities $a_{0,1}=1.3$, $a_{0,2}=1.0$, and fixed angle $\theta_{k'}\approx2.0$ mrad. (a) $\nu=2$ ($\omega_2=2\omega_1$); (b) $\nu=3$ ($\omega_2=3\omega_1$). At photon energies where two multiphoton channels become degenerate, the spectrum shows simultaneous population of two OAM modes, forming vortex $\gamma$ photon in superposition states obeying $|\Delta\ell'|=\nu-1$ (equal helicities). Insets: transverse intensity and phase distributions at the harmonic peak (vertical line). Other parameters are given in the text; insets in (b) share the same scales and axes as (a).}
\label{fig_2}
\end{figure}
\begin{figure}[!t]
		\setlength{\abovecaptionskip}{-0.0cm}
		\includegraphics[width=0.995\linewidth]{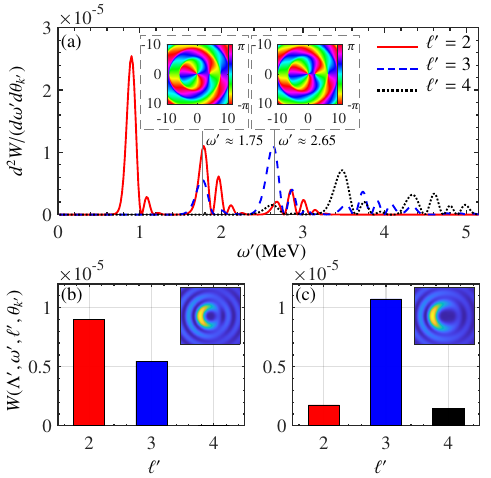}
		\begin{picture}(300,0)
		\end{picture}
\setlength{\abovecaptionskip}{-0.5 cm}
\caption{Generation of vortex $\gamma$ photon in superposition states in a three-frequency laser field. Parameters: $\Lambda'=-1$, $a_{0,1}=1.4$, $a_{0,2}=1.2$, $a_{0,3}=1.0$, $\omega_2=2\omega_1$, $\omega_3=3\omega_1$. (a) Energy spectrum at $\theta_{k'}\approx2.4\,\mathrm{mrad}$. Degenerate multiphoton channels at different photon energies produce distinct superposition manifolds; two vertical lines mark representative degeneracy points, with phase profiles shown in the insets. (b,c) OAM-mode compositions and transverse intensity distributions for the two superposition states marked in (a). Intensity and phase panels follow the conventions of Fig.~\ref{fig_2} (same color scales, axes, and plotted quantities).}
\label{fig_3}
\end{figure}
\begin{figure}[!t]
		\setlength{\abovecaptionskip}{-0.0cm}
		\includegraphics[width=0.995\linewidth]{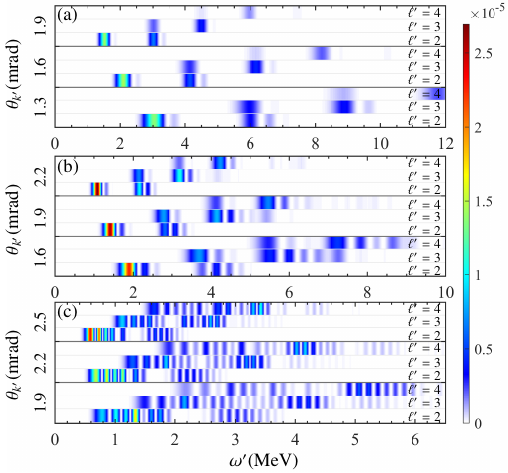}
		\begin{picture}(300,0)
		\end{picture}
\setlength{\abovecaptionskip}{-0.5 cm}
\caption{Effect of laser intensity on the formation of vortex $\gamma$ photon in superposition states for the two-frequency configuration and $\omega_2=2\omega_1$.
(a) $a_{0,1}=0.8$, $a_{0,2}=0.5$;
(b) $a_{0,1}=1.3$, $a_{0,2}=1.0$;
(c) $a_{0,1}=3.3$, $a_{0,2}=3.0$.
The emitted photon helicity is $\Lambda'=-1$.
Each subplot shows the probability of vortex $\gamma$ photon generation for different OAM values at three $\theta_{k'}$ values.}
\label{fig_4}
\end{figure}

In this Letter, we show that vortex $\gamma$ photon in superposition states can be tunably generated via NCS driven by multifrequency CP lasers (Fig.~\ref{fig_1}). Therefore, such photon can serve as a novel physical resource for coherent quantum-state manipulation in the $\gamma$-ray regime. The underlying mechanism is quantum interference between multiphoton absorption pathways that are degenerate in energy--momentum but distinct in total angular momentum (TAM). For two-frequency drivers, we derive a universal relation: the OAM separation satisfies $\Delta\ell'=\nu-1$ for equal laser helicities and $\Delta\ell'=\nu+1$ for opposite helicities, such that the frequency ratio $\nu$ fixes the mode spacing while the intensities $(a_{0,1},a_{0,2})$ control the modal weights (Fig.~\ref{fig_2}). This framework also extends to three-frequency fields, offering additional degrees of freedom for generating diverse superposition modes across the spectrum (Fig.~\ref{fig_3}). At higher intensities, ponderomotive broadening causes the interfering channels to merge into continuous bands of structured radiation (Fig.~\ref{fig_4}). The generated $\Delta\ell'$-dependent interference patterns provide target-independent fingerprints of photon spatial structure even in plane-wave environments where absolute OAM is difficult to resolve. These results establish vortex $\gamma$ photon in superposition states as a new resource for coherent quantum-state engineering at the $\gamma$-ray frontier. Throughout, natural units are used ($\hbar = c = 1$), the electron charge $e = -|e|$, and the electron mass is denoted as $m_e$.

\textit{Control Mechanism.—}We consider the NCS of an ultrarelativistic electron with initial four‑momentum $p^\mu=(\varepsilon,\bm{p})$, interacting with a counterpropagating multifrequency CP laser pulse $A^\mu(\phi)=\sum_{j=1}^{N_c} A_j\, g_j(\phi_j)\operatorname{Re}\bigl[(0,1,-i\Lambda_j,0)\,e^{i\phi_j}\bigr]$. Here $\phi_j=k_j\!\cdot x$ is the laser phase, with laser-photon four-momentum $k_j^\mu=\omega_j(1,0,0,-1)$, $g_j(\phi_j)$ is the pulse-shape function, and the $j$th mode has frequency $\omega_j=\nu_j\omega_1$, normalized intensity parameter $a_{0,j}=eA_j/m_e$, and helicity $\Lambda_j=\pm1$. The electron is scattered to a final state with four-momentum $p'^\mu=(\varepsilon',\bm{p}')$, emitting a $\gamma$ photon with four-momentum $k'^\mu=(\omega',\bm{k}')$. In the Furry picture, the NCS amplitude is defined by $S_{fi}=-ie\int d^4x\,\bar{\Psi}_{p'}(x)\gamma^\mu\Psi_p(x)A'_\mu(x)$, where $\Psi_p$ and $\Psi_{p'}$ are Volkov electron states and $A'_\mu$ is the emitted photon state \cite{Ritus:1985vta,DiPiazza2012,Fedotov2023}. For a multifrequency laser, this amplitude reduces to a coherent sum over multiphoton absorption channels $\bm{n}=(n_1,\dots,n_{N_c})$, expressed as $S_{fi}=ie\,(2\pi)^3/\omega_1\,\delta^{(3)}(\bm{p}+s\,\bm{k}_1-\bm{p}'-\bm{k}')\sum_{\bm{n}}\mathcal{M}_{\bm{n}}(s)e^{i\Phi_{\bm{n}}}$, where $s$ is the continuous light-front momentum fraction, $\Phi_{\bm{n}}$ is the channel-dependent phase, and $\mathcal{M}_{\bm{n}}(s)$ is the channel amplitude [see Supplemental Material (SM), Secs. I A and B \cite{supplemental}].  

To characterize the emitted structured photon, we project the final state onto a Bessel mode $|k'_z, k'_\perp, m', \Lambda'\rangle$ defined by its longitudinal momentum $k'_z$, transverse-momentum modulus $k'_\perp = \omega' \sin\theta_{k'}$ (where $\theta_{k'}$ is the vortex cone angle), TAM projection $m'$, and helicity $\Lambda'$ \cite{Jentschura:2010ap,jentschura2011compton}. In the paraxial regime ($\theta_{k'} \sim 1/\gamma$, where $\gamma = \varepsilon/m_e$ is the initial electron Lorentz factor), spin and OAM are separable. The intrinsic OAM then satisfies $\ell' = m' - \Lambda' = \sum n_j \Lambda_j + \lambda - \lambda' - \Lambda'$, where $n_j$ denotes the number of photons absorbed from the $j$th laser mode, and $\lambda$ ($\lambda'$) represents the initial (final) electron spin. For moderate laser intensities ($a_{0,j} \sim 1$), where no-spin-flip channels ($\lambda = \lambda'$) dominate, this reduces to the selection rule $\ell' = \sum n_j \Lambda_j - \Lambda'$. Consequently, the emitted radiation is a vortex-$\gamma$ photon state described by a coherent superposition of OAM modes with identical kinematics and helicity, $|\Psi_\gamma\rangle = \sum_{\ell'} c_{\ell'} | \ell',\Lambda'\rangle$. Here, the complex coefficients $c_{\ell'}$ are determined by the multiphoton transition amplitudes $\mathcal{M}_{\bm{n}}$ (see SM, Sec.~I C \cite{supplemental}). 

In pulsed fields, each multiphoton channel $\bm{n}$ undergoes intensity-dependent frequency broadening. Defining the effective harmonic index $N(\bm{n}) = \sum_j \nu_j n_j$, the spectral support of a channel is approximately $N(\bm{n}) + \beta_\Sigma \lesssim s \lesssim N(\bm{n})$, where $\beta_\Sigma = \sum_j \left( \frac{1}{2k_1\cdot p} - \frac{1}{2k_1\cdot p'} \right) m_e^2 a^2_{0,j}$ is the ponderomotive phase shift. Two distinct channels $\bm{n}$ and $\bm{n}'$ contribute to the same photon kinematics whenever their support intervals overlap, $|N(\bm{n}) - N(\bm{n}')| \lesssim |\beta_\Sigma|$. In these overlap regions, the emitted photon is produced as a coherent superposition of distinct OAM components, with modal amplitudes and relative phases determined by the complex channel amplitudes $\mathcal{M}_{\bm{n}}$ and $\mathcal{M}_{\bm{n}'}$.

For a two-frequency laser with ratio $\nu \equiv \omega_2/\omega_1$, exact channel degeneracies arise because $k_2 = \nu k_1$. Channel pairs $(n_1, n_2)$ and $(n'_1, n'_2) = (n_1 - \nu, n_2 + 1)$ satisfy $N(\bm{n}) = N(\bm{n}')$, ensuring maximal spectral overlap. The resulting OAM separation, $\Delta\ell' = |(n_1\Lambda_1 + n_2\Lambda_2) - (n'_1\Lambda_1 + n'_2\Lambda_2)|$, obeys a universal relation:
\begin{equation}
|\Delta \ell'| =
\begin{cases}
\nu - 1, & \text{Equal helicities } (\Lambda_1=\Lambda_2),\\
\nu + 1, & \text{Opposite helicities } (\Lambda_1=-\Lambda_2).
\end{cases}
\end{equation}
The frequency ratio thus fixes the accessible OAM separation, while the emission kinematics and pulse broadening determine where in the spectrum the superposition is realized. 

We evaluate the vortex-resolved emission probability by projecting the final state onto vortex modes:
\begin{eqnarray}
\begin{aligned}
\frac{d^2W}{d\omega^\prime d\theta_{k^\prime}}
=\frac{e^2}{4\pi}\frac{k^\prime_\perp}{(k_1p)(k_1p^\prime)}
\big|\sum_{\bm{n}}\delta_{\ell^\prime,\,m'_{\bm{n}}-\Lambda'} (-i)^{m'_{\bm{n}}}
\mathcal{M}_{\bm{n}}\big|^2 \, ,
\end{aligned}
\label{rate_vortex}
\end{eqnarray}
where $m'_{\bm{n}}={\sum_j n_j \Lambda_j}$ is the TAM contribution due to multiphoton absorption. While Eq.~\eqref{rate_vortex} gives the probability of emission into specific OAM eigenmodes $|\ell',\Lambda'\rangle$, the photon is emitted at the vertex as a coherent superposition. The Kronecker delta $\delta_{\ell^\prime,\,m'_{\bm{n}}-\Lambda'}$ acts as a projection resolving the modal weights $|c_{\ell'}|^2$. 
In the overlap regions, the amplitudes $\mathcal{M}_{\bm{n}}$ ensure that the emitted state is not merely an incoherent mixture: the relative phases between the contributing channels determine a well-defined spatial structure, manifesting in the robust $|\Delta\ell'|$-fold azimuthal interference patterns and phase singularities shown below---a direct signature of the internal quantum structure of the emitted photon.

\textit{Manipulated Generation.—}
We present numerical results illustrating the generation of vortex $\gamma$ photon in superposition states by evaluating the emission rate via Eq.~\eqref{rate_vortex}. We consider a head-on collision between a $1$ GeV electron beam and a counterpropagating, multifrequency, corotating (equal-helicity) CP laser pulse with moderate intensity $a_{0,j} \sim 1$, where spectral overlap occurs. The pulse shape is $g_j(\phi_j) = \cos^2\!\left( \phi_j / (2\nu_j N_{\text{cycle}}) \right)$ with $N_{\text{cycle}} = 10$, and the fundamental laser frequency is $\omega_1 = 1.55$eV. 

Figure~\ref{fig_2} shows the energy-resolved emission probability at a fixed vortex cone angle $\theta_{k'}$ for two-frequency CP laser fields with $\omega_2=\nu\omega_1$ ($\nu=2,3$). Distinct harmonic peaks corresponding to vortex $\gamma$ photon with different intrinsic OAM are clearly resolved. At specific photon energies, the spectrum simultaneously populates two vortex modes. This signals the overlap of kinematically degenerate multiphoton channels: for $\nu=2$, the interference between the $(n_1,n_2)=(2,0)$ and $(0,1)$ pathways yields a coherent superposition of intrinsic OAM modes $|\Psi_\gamma\rangle=c_1|\ell'=2\rangle+c_2|\ell'=3\rangle$.

The insets show the corresponding transverse intensity and phase profiles. For a two-mode superposition $c_1|\ell'_1\rangle + c_2|\ell'_2\rangle$, the transverse intensity follows $I(\varphi)\propto|c_1 e^{i\ell'_1\varphi} + c_2 e^{i\ell'_2\varphi}|^2 = |c_1|^2 + |c_2|^2 + 2|c_1 c_2|\cos(\Delta\ell'\varphi + \delta)$, where $\Delta\ell'=|\ell'_1-\ell'_2|$. The cosine term produces $\Delta\ell'$-fold azimuthal dark notches (their angular positions depend on $\delta$), while the squared Bessel function ensures a dominant main ring in the radial direction. This directly accounts for the multi-lobed intensity patterns and phase discontinuities seen in the figure.

Increasing the frequency ratio to $\nu=3$ [Fig.~\ref{fig_2}(b)] activates the $(3,0)$ and $(0,1)$ channel pair, yielding superpositions with $\Delta\ell'=2$. This directly verifies the universal relation $\Delta\ell'=\nu-1$ and demonstrates a clear departure from single-frequency driving, where only adjacent harmonics ($\Delta\ell'=1$) can interfere. These interference structures constitute a robust, target-independent fingerprint of OAM superposition, with the OAM separation set by the laser frequency ratio. The opposite-helicity case, obeying the $\Delta\ell'=\nu+1$ branch, is verified in Fig.~S1 of the SM, where representative spectra, mode decompositions, and transverse profiles are presented \cite{supplemental}.
	
Adding a third frequency expands the accessible OAM superposition states. Figure~\ref{fig_3}(a) shows results for a three-frequency driver $(\omega_1,2\omega_1,3\omega_1)$. At a fixed emission angle, the spectrum exhibits multiple overlapping harmonic peaks, indicating the activation of several distinct degeneracy manifolds. The superposition type depends on photon energy $\omega'$: at $1.75$ MeV, the $\ell'=2$ and $\ell'=3$ modes overlap, while at $2.65$ MeV, the coexistence of the $3\omega_1$ component and mixed-frequency pathways enables a three-mode superposition involving $\ell'=2,3,4$. As seen in Fig.~\ref{fig_3}(c), the $\ell'=3$ mode dominates, so the interference effectively reduces to a two‑mode case with $\Delta\ell'=1$, producing a single dark notch whose azimuthal position is set by the relative phases.

Unlike the two-frequency case (where a single ratio $\nu$ fixes a global $\Delta\ell'$), the multichromatic field enables the simultaneous synthesis of spectrally separated states with different modal compositions. The phase profiles shown in the insets of Fig.~\ref{fig_3}(a), together with the mode decompositions and transverse intensity distributions in Figs.~\ref{fig_3}(b) and \ref{fig_3}(c), provide direct signatures of coherent vortex-mode superposition. Because these interference structures originate from intrinsic quantum-state properties rather than classical beam shaping, they remain observable even in plane-wave scattering environments, ensuring the scheme’s utility for angular-momentum-resolved spectroscopy.

The role of laser intensity in defining the operational regimes of our mechanism is illustrated in Fig.~\ref{fig_4}. At weak intensities [Fig.~\ref{fig_4}(a)], the spectrum consists of isolated harmonic lines where superposition occurs only at specific energies due to minimal channel overlap. As the intensity increases [Figs.~\ref{fig_4}(b) and (c)], the ponderomotive broadening parameter $|\beta_\Sigma|$ grows, allowing the overlap criterion to be satisfied more readily. Consequently, discrete lines broaden and coalesce into continuous bands. This marks the transition into a strong‑field regime where a dense manifold of interfering multiphoton pathways contributes with comparable weight, significantly increasing spectral complexity and the richness of accessible structured photon states.  

Accompanying this transition is a systematic broadening of the emission cone and a redshift of the spectral weight. The angular expansion follows  $\theta \sim a_{\text{eff}}/\gamma$ (equivalently $\theta \sim 1/\gamma^*$). This behavior stems from the laser-induced increase of the electron's effective mass $m^* = m_e\sqrt{1 + a_{0,1}^2 + a_{0,2}^2}$, which suppresses the longitudinal Lorentz factor $\gamma^* = \varepsilon/m^*$ and thus widens the radiation aperture. Concurrently, the spectral range shifts toward lower photon energies due to the intensity-dependent reduction of the kinematic energy cutoff, which for the dominant channels scales as $\delta_n \propto 1/(1 + a_{0,1}^2 + a_{0,2}^2)$. This redshift reflects the increased inertia of the dressed electron and the intensified recoil effects as the interaction enters the deep nonlinear regime. 
A corresponding intensity-driven transition is found for the opposite-helicity configuration; representative results are shown in Fig.~S2 of the SM \cite{supplemental}. 

In plane-wave scattering geometries, absolute OAM eigenvalues are generally difficult to infer directly. Coherent vortex superpositions, by contrast, can imprint directly observable azimuthal interference in the measured signal \cite{PhysRevA.92.013401,Ababekri:2024cyd}. The coherence between distinct OAM components acts as an intrinsic azimuthal modulation of the scattering pattern, which can be represented by a factor $F(\varphi_{k'})=1+\cos(\Delta\ell'\,\varphi_{k'}+\delta)$, thereby imprinting a $|\Delta\ell'|$-fold symmetry on the radiation cone. Because this self-interference is generated coherently at the emission vertex, it provides a target-independent fingerprint that is robust against beam jitter and alignment errors, while relaxing the stringent centering requirements associated with single-mode probes. NCS driven by multifrequency lasers therefore offers a practical and robust route to angular-momentum-resolved spectroscopy at the $\gamma$-ray frontier.

In summary, we have shown that NCS driven by multifrequency lasers enables the direct generation of vortex $\gamma$ photon in superposition states through interference between energy-degenerate multiphoton channels of distinct TAM. 
The OAM separation $\Delta\ell'$ is fixed by the laser frequency ratio, while the relative modal weights are tuned by the laser intensities, providing state control inaccessible with single-frequency drivers. The resulting $\Delta\ell'$-fold interference patterns furnish robust, target-independent fingerprints of $\gamma$ photon spatial structure, ensuring that the OAM degree of freedom remains accessible even at ultrahigh energies. Because the mechanism is formulated at the amplitude level within strong-field QED, it provides a natural starting point for extensions to structured initial electron states, including vortex and wave-packet electrons \cite{bliokh2017theory,Wong2021}, thereby further connecting structured-light physics, strong-field QED, and nuclear photonics.\\

{\it Acknowledgments---}The work is supported by the National Natural Science Foundation of China (Grants No. 12425510, No. U2267204, No. 12441506, No. 12505276), the National Key Research and Development (R\&D) Program (Grant No. 2024YFA1610900), the Science Challenge Project (No. TZ2025012), and the Innovative Scientific Program of CNNC.

\bibliography{newmanuscript}

\begin{thebibliography}{68}%
\makeatletter
\providecommand \@ifxundefined [1]{%
 \@ifx{#1\undefined}
}%
\providecommand \@ifnum [1]{%
 \ifnum #1\expandafter \@firstoftwo
 \else \expandafter \@secondoftwo
 \fi
}%
\providecommand \@ifx [1]{%
 \ifx #1\expandafter \@firstoftwo
 \else \expandafter \@secondoftwo
 \fi
}%
\providecommand \natexlab [1]{#1}%
\providecommand \enquote  [1]{``#1''}%
\providecommand \bibnamefont  [1]{#1}%
\providecommand \bibfnamefont [1]{#1}%
\providecommand \citenamefont [1]{#1}%
\providecommand \href@noop [0]{\@secondoftwo}%
\providecommand \href [0]{\begingroup \@sanitize@url \@href}%
\providecommand \@href[1]{\@@startlink{#1}\@@href}%
\providecommand \@@href[1]{\endgroup#1\@@endlink}%
\providecommand \@sanitize@url [0]{\catcode `\\12\catcode `\$12\catcode
  `\&12\catcode `\#12\catcode `\^12\catcode `\_12\catcode `\%12\relax}%
\providecommand \@@startlink[1]{}%
\providecommand \@@endlink[0]{}%
\providecommand \url  [0]{\begingroup\@sanitize@url \@url }%
\providecommand \@url [1]{\endgroup\@href {#1}{\urlprefix }}%
\providecommand \urlprefix  [0]{URL }%
\providecommand \Eprint [0]{\href }%
\providecommand \doibase [0]{https://doi.org/}%
\providecommand \selectlanguage [0]{\@gobble}%
\providecommand \bibinfo  [0]{\@secondoftwo}%
\providecommand \bibfield  [0]{\@secondoftwo}%
\providecommand \translation [1]{[#1]}%
\providecommand \BibitemOpen [0]{}%
\providecommand \bibitemStop [0]{}%
\providecommand \bibitemNoStop [0]{.\EOS\space}%
\providecommand \EOS [0]{\spacefactor3000\relax}%
\providecommand \BibitemShut  [1]{\csname bibitem#1\endcsname}%
\let\auto@bib@innerbib\@empty
\bibitem [{\citenamefont {Allen}\ \emph {et~al.}(1992)\citenamefont {Allen},
  \citenamefont {Beijersbergen}, \citenamefont {Spreeuw},\ and\ \citenamefont
  {Woerdman}}]{Allen:1992zz}%
  \BibitemOpen
  \bibfield  {author} {\bibinfo {author} {\bibfnamefont {L.}~\bibnamefont
  {Allen}}, \bibinfo {author} {\bibfnamefont {M.~W.}\ \bibnamefont
  {Beijersbergen}}, \bibinfo {author} {\bibfnamefont {R.~J.~C.}\ \bibnamefont
  {Spreeuw}},\ and\ \bibinfo {author} {\bibfnamefont {J.~P.}\ \bibnamefont
  {Woerdman}},\ }\bibfield  {title} {\bibinfo {title} {Orbital angular momentum
  of light and the transformation of laguerre-gaussian laser modes},\ }\href
  {https://doi.org/10.1103/PhysRevA.45.8185} {\bibfield  {journal} {\bibinfo
  {journal} {Phys. Rev. A}\ }\textbf {\bibinfo {volume} {45}},\ \bibinfo
  {pages} {8185} (\bibinfo {year} {1992})}\BibitemShut {NoStop}%
\bibitem [{\citenamefont {Shen}\ \emph {et~al.}(2019)\citenamefont {Shen},
  \citenamefont {Wang}, \citenamefont {Xie}, \citenamefont {Min}, \citenamefont
  {Fu}, \citenamefont {Liu}, \citenamefont {Gong},\ and\ \citenamefont
  {Yuan}}]{Shen2019}%
  \BibitemOpen
  \bibfield  {author} {\bibinfo {author} {\bibfnamefont {Y.}~\bibnamefont
  {Shen}}, \bibinfo {author} {\bibfnamefont {X.}~\bibnamefont {Wang}}, \bibinfo
  {author} {\bibfnamefont {Z.}~\bibnamefont {Xie}}, \bibinfo {author}
  {\bibfnamefont {C.}~\bibnamefont {Min}}, \bibinfo {author} {\bibfnamefont
  {X.}~\bibnamefont {Fu}}, \bibinfo {author} {\bibfnamefont {Q.}~\bibnamefont
  {Liu}}, \bibinfo {author} {\bibfnamefont {M.}~\bibnamefont {Gong}},\ and\
  \bibinfo {author} {\bibfnamefont {X.}~\bibnamefont {Yuan}},\ }\bibfield
  {title} {\bibinfo {title} {Optical vortices 30 years on: Oam manipulation
  from topological charge to multiple singularities},\ }\href
  {https://doi.org/10.1038/s41377-019-0194-2} {\bibfield  {journal} {\bibinfo
  {journal} {Light Sci. Appl.}\ }\textbf {\bibinfo {volume} {8}},\ \bibinfo
  {pages} {90} (\bibinfo {year} {2019})}\BibitemShut {NoStop}%
\bibitem [{\citenamefont {Forbes}\ \emph {et~al.}(2021)\citenamefont {Forbes},
  \citenamefont {de~Oliveira},\ and\ \citenamefont {Dennis}}]{Forbes:2021tpp}%
  \BibitemOpen
  \bibfield  {author} {\bibinfo {author} {\bibfnamefont {A.}~\bibnamefont
  {Forbes}}, \bibinfo {author} {\bibfnamefont {M.}~\bibnamefont
  {de~Oliveira}},\ and\ \bibinfo {author} {\bibfnamefont {M.~R.}\ \bibnamefont
  {Dennis}},\ }\bibfield  {title} {\bibinfo {title} {Structured light},\ }\href
  {https://doi.org/10.1038/s41566-021-00780-4} {\bibfield  {journal} {\bibinfo
  {journal} {Nat. Photonics}\ }\textbf {\bibinfo {volume} {15}},\ \bibinfo
  {pages} {253} (\bibinfo {year} {2021})}\BibitemShut {NoStop}%
\bibitem [{\citenamefont {Forbes}\ \emph {et~al.}(2025)\citenamefont {Forbes},
  \citenamefont {Nothlawala},\ and\ \citenamefont
  {Vallés}}]{forbes2025progress}%
  \BibitemOpen
  \bibfield  {author} {\bibinfo {author} {\bibfnamefont {A.}~\bibnamefont
  {Forbes}}, \bibinfo {author} {\bibfnamefont {F.}~\bibnamefont {Nothlawala}},\
  and\ \bibinfo {author} {\bibfnamefont {A.}~\bibnamefont {Vallés}},\
  }\bibfield  {title} {\bibinfo {title} {Progress in quantum structured
  light},\ }\href {https://doi.org/10.1038/s41566-025-01795-x} {\bibfield
  {journal} {\bibinfo  {journal} {Nat. Photonics}\ }\textbf {\bibinfo {volume}
  {19}},\ \bibinfo {pages} {1291} (\bibinfo {year} {2025})}\BibitemShut
  {NoStop}%
\bibitem [{\citenamefont {Mair}\ \emph {et~al.}(2001)\citenamefont {Mair},
  \citenamefont {Weihs},\ and\ \citenamefont
  {Zeilinger}}]{mairEntanglementOrbitalAngular2001}%
  \BibitemOpen
  \bibfield  {author} {\bibinfo {author} {\bibfnamefont {A.}~\bibnamefont
  {Mair}}, \bibinfo {author} {\bibfnamefont {G.}~\bibnamefont {Weihs}},\ and\
  \bibinfo {author} {\bibfnamefont {A.}~\bibnamefont {Zeilinger}},\ }\bibfield
  {title} {\bibinfo {title} {Entanglement of the orbital angular momentum
  states of photons},\ }\href {https://doi.org/10.1038/35085529} {\bibfield
  {journal} {\bibinfo  {journal} {Nature}\ }\textbf {\bibinfo {volume} {412}},\
  \bibinfo {pages} {313} (\bibinfo {year} {2001})}\BibitemShut {NoStop}%
\bibitem [{\citenamefont {Vaziri}\ \emph {et~al.}(2002)\citenamefont {Vaziri},
  \citenamefont {Weihs},\ and\ \citenamefont {Zeilinger}}]{Vaziri_2002}%
  \BibitemOpen
  \bibfield  {author} {\bibinfo {author} {\bibfnamefont {A.}~\bibnamefont
  {Vaziri}}, \bibinfo {author} {\bibfnamefont {G.}~\bibnamefont {Weihs}},\ and\
  \bibinfo {author} {\bibfnamefont {A.}~\bibnamefont {Zeilinger}},\ }\bibfield
  {title} {\bibinfo {title} {Superpositions of the orbital angular momentum for
  applications in quantum experiments},\ }\href
  {https://doi.org/10.1088/1464-4266/4/2/367} {\bibfield  {journal} {\bibinfo
  {journal} {J. Opt. B: Quantum Semiclass. Opt.}\ }\textbf {\bibinfo {volume}
  {4}},\ \bibinfo {pages} {S47} (\bibinfo {year} {2002})}\BibitemShut {NoStop}%
\bibitem [{\citenamefont {Molina-Terriza}\ \emph {et~al.}(2007)\citenamefont
  {Molina-Terriza}, \citenamefont {Torres},\ and\ \citenamefont
  {Torner}}]{MolinaTerriza2007}%
  \BibitemOpen
  \bibfield  {author} {\bibinfo {author} {\bibfnamefont {G.}~\bibnamefont
  {Molina-Terriza}}, \bibinfo {author} {\bibfnamefont {J.~P.}\ \bibnamefont
  {Torres}},\ and\ \bibinfo {author} {\bibfnamefont {L.}~\bibnamefont
  {Torner}},\ }\bibfield  {title} {\bibinfo {title} {Twisted photons},\ }\href
  {https://doi.org/10.1038/nphys607} {\bibfield  {journal} {\bibinfo  {journal}
  {Nat. Phys.}\ }\textbf {\bibinfo {volume} {3}},\ \bibinfo {pages} {305}
  (\bibinfo {year} {2007})}\BibitemShut {NoStop}%
\bibitem [{\citenamefont {Yao}\ and\ \citenamefont {Padgett}(2011)}]{Yao2011}%
  \BibitemOpen
  \bibfield  {author} {\bibinfo {author} {\bibfnamefont {A.~M.}\ \bibnamefont
  {Yao}}\ and\ \bibinfo {author} {\bibfnamefont {M.~J.}\ \bibnamefont
  {Padgett}},\ }\bibfield  {title} {\bibinfo {title} {Orbital angular momentum:
  origins, behavior and applications},\ }\href
  {https://doi.org/10.1364/AOP.3.000161} {\bibfield  {journal} {\bibinfo
  {journal} {Adv. Opt. Photon.}\ }\textbf {\bibinfo {volume} {3}},\ \bibinfo
  {pages} {161} (\bibinfo {year} {2011})}\BibitemShut {NoStop}%
\bibitem [{\citenamefont {Erhard}\ \emph {et~al.}(2020)\citenamefont {Erhard},
  \citenamefont {Krenn},\ and\ \citenamefont {Zeilinger}}]{erhard2020advances}%
  \BibitemOpen
  \bibfield  {author} {\bibinfo {author} {\bibfnamefont {M.}~\bibnamefont
  {Erhard}}, \bibinfo {author} {\bibfnamefont {M.}~\bibnamefont {Krenn}},\ and\
  \bibinfo {author} {\bibfnamefont {A.}~\bibnamefont {Zeilinger}},\ }\bibfield
  {title} {\bibinfo {title} {Advances in high-dimensional quantum
  entanglement},\ }\href {https://doi.org/10.1038/s42254-020-0193-5} {\bibfield
   {journal} {\bibinfo  {journal} {Nat. Rev. Phys.}\ }\textbf {\bibinfo
  {volume} {2}},\ \bibinfo {pages} {365} (\bibinfo {year} {2020})}\BibitemShut
  {NoStop}%
\bibitem [{\citenamefont {Kapale}\ and\ \citenamefont
  {Dowling}(2005)}]{kapaleVortexPhaseQubit2005}%
  \BibitemOpen
  \bibfield  {author} {\bibinfo {author} {\bibfnamefont {K.~T.}\ \bibnamefont
  {Kapale}}\ and\ \bibinfo {author} {\bibfnamefont {J.~P.}\ \bibnamefont
  {Dowling}},\ }\bibfield  {title} {\bibinfo {title} {Vortex phase qubit:
  Generating arbitrary, counterrotating, coherent superpositions in
  bose-einstein condensates via optical angular momentum beams},\ }\href
  {https://doi.org/10.1103/PhysRevLett.95.173601} {\bibfield  {journal}
  {\bibinfo  {journal} {Phys. Rev. Lett.}\ }\textbf {\bibinfo {volume} {95}},\
  \bibinfo {pages} {173601} (\bibinfo {year} {2005})}\BibitemShut {NoStop}%
\bibitem [{\citenamefont {Moxley}\ \emph {et~al.}(2016)\citenamefont {Moxley},
  \citenamefont {Dowling}, \citenamefont {Dai},\ and\ \citenamefont
  {Byrnes}}]{Moxley:2015yil}%
  \BibitemOpen
  \bibfield  {author} {\bibinfo {author} {\bibfnamefont {F.~I.}\ \bibnamefont
  {Moxley}}, \bibinfo {author} {\bibfnamefont {J.~P.}\ \bibnamefont {Dowling}},
  \bibinfo {author} {\bibfnamefont {W.}~\bibnamefont {Dai}},\ and\ \bibinfo
  {author} {\bibfnamefont {T.}~\bibnamefont {Byrnes}},\ }\bibfield  {title}
  {\bibinfo {title} {Sagnac interferometry with coherent vortex superposition
  states in exciton-polariton condensates},\ }\href
  {https://doi.org/10.1103/PhysRevA.93.053603} {\bibfield  {journal} {\bibinfo
  {journal} {Phys. Rev. A}\ }\textbf {\bibinfo {volume} {93}},\ \bibinfo
  {pages} {053603} (\bibinfo {year} {2016})}\BibitemShut {NoStop}%
\bibitem [{\citenamefont {Babazadeh}\ \emph {et~al.}(2017)\citenamefont
  {Babazadeh}, \citenamefont {Erhard}, \citenamefont {Wang}, \citenamefont
  {Malik}, \citenamefont {Nouroozi}, \citenamefont {Krenn},\ and\ \citenamefont
  {Zeilinger}}]{Babazadeh:2017ioi}%
  \BibitemOpen
  \bibfield  {author} {\bibinfo {author} {\bibfnamefont {A.}~\bibnamefont
  {Babazadeh}}, \bibinfo {author} {\bibfnamefont {M.}~\bibnamefont {Erhard}},
  \bibinfo {author} {\bibfnamefont {F.}~\bibnamefont {Wang}}, \bibinfo {author}
  {\bibfnamefont {M.}~\bibnamefont {Malik}}, \bibinfo {author} {\bibfnamefont
  {R.}~\bibnamefont {Nouroozi}}, \bibinfo {author} {\bibfnamefont
  {M.}~\bibnamefont {Krenn}},\ and\ \bibinfo {author} {\bibfnamefont
  {A.}~\bibnamefont {Zeilinger}},\ }\bibfield  {title} {\bibinfo {title}
  {High-dimensional single-photon quantum gates: Concepts and experiments},\
  }\href {https://doi.org/10.1103/PhysRevLett.119.180510} {\bibfield  {journal}
  {\bibinfo  {journal} {Phys. Rev. Lett.}\ }\textbf {\bibinfo {volume} {119}},\
  \bibinfo {pages} {180510} (\bibinfo {year} {2017})}\BibitemShut {NoStop}%
\bibitem [{\citenamefont {Huang}\ \emph {et~al.}(2025)\citenamefont {Huang},
  \citenamefont {Mao}, \citenamefont {Li}, \citenamefont {Yuan}, \citenamefont
  {Zheng}, \citenamefont {Zhai}, \citenamefont {Dai}, \citenamefont {Fu},
  \citenamefont {Bao}, \citenamefont {Yang} \emph {et~al.}}]{Huang:2025jqa}%
  \BibitemOpen
  \bibfield  {author} {\bibinfo {author} {\bibfnamefont {J.}~\bibnamefont
  {Huang}}, \bibinfo {author} {\bibfnamefont {J.}~\bibnamefont {Mao}}, \bibinfo
  {author} {\bibfnamefont {X.}~\bibnamefont {Li}}, \bibinfo {author}
  {\bibfnamefont {J.}~\bibnamefont {Yuan}}, \bibinfo {author} {\bibfnamefont
  {Y.}~\bibnamefont {Zheng}}, \bibinfo {author} {\bibfnamefont
  {C.}~\bibnamefont {Zhai}}, \bibinfo {author} {\bibfnamefont {T.}~\bibnamefont
  {Dai}}, \bibinfo {author} {\bibfnamefont {Z.}~\bibnamefont {Fu}}, \bibinfo
  {author} {\bibfnamefont {J.}~\bibnamefont {Bao}}, \bibinfo {author}
  {\bibfnamefont {Y.}~\bibnamefont {Yang}}, \emph {et~al.},\ }\bibfield
  {title} {\bibinfo {title} {Integrated optical entangled quantum vortex
  emitters},\ }\href {https://doi.org/10.1038/s41566-025-01620-5} {\bibfield
  {journal} {\bibinfo  {journal} {Nat. Photonics}\ }\textbf {\bibinfo {volume}
  {19}},\ \bibinfo {pages} {471} (\bibinfo {year} {2025})}\BibitemShut
  {NoStop}%
\bibitem [{\citenamefont {Lu}\ \emph {et~al.}(2023)\citenamefont {Lu},
  \citenamefont {Guo}, \citenamefont {Li}, \citenamefont {Ababekri},
  \citenamefont {Chen}, \citenamefont {Fu}, \citenamefont {Lv}, \citenamefont
  {Xu}, \citenamefont {Kong}, \citenamefont {Niu} \emph {et~al.}}]{Lu:2023wrf}%
  \BibitemOpen
  \bibfield  {author} {\bibinfo {author} {\bibfnamefont {Z.-W.}\ \bibnamefont
  {Lu}}, \bibinfo {author} {\bibfnamefont {L.}~\bibnamefont {Guo}}, \bibinfo
  {author} {\bibfnamefont {Z.-Z.}\ \bibnamefont {Li}}, \bibinfo {author}
  {\bibfnamefont {M.}~\bibnamefont {Ababekri}}, \bibinfo {author}
  {\bibfnamefont {F.-Q.}\ \bibnamefont {Chen}}, \bibinfo {author}
  {\bibfnamefont {C.}~\bibnamefont {Fu}}, \bibinfo {author} {\bibfnamefont
  {C.}~\bibnamefont {Lv}}, \bibinfo {author} {\bibfnamefont {R.}~\bibnamefont
  {Xu}}, \bibinfo {author} {\bibfnamefont {X.}~\bibnamefont {Kong}}, \bibinfo
  {author} {\bibfnamefont {Y.-F.}\ \bibnamefont {Niu}}, \emph {et~al.},\
  }\bibfield  {title} {\bibinfo {title} {Manipulation of giant multipole
  resonances via vortex $\ensuremath{\gamma}$ photons},\ }\href
  {https://doi.org/10.1103/PhysRevLett.131.202502} {\bibfield  {journal}
  {\bibinfo  {journal} {Phys. Rev. Lett.}\ }\textbf {\bibinfo {volume} {131}},\
  \bibinfo {pages} {202502} (\bibinfo {year} {2023})}\BibitemShut {NoStop}%
\bibitem [{\citenamefont {Ujeniuc}\ and\ \citenamefont
  {Suvaila}(2024)}]{Ujeniuc:2024hse}%
  \BibitemOpen
  \bibfield  {author} {\bibinfo {author} {\bibfnamefont {S.}~\bibnamefont
  {Ujeniuc}}\ and\ \bibinfo {author} {\bibfnamefont {R.}~\bibnamefont
  {Suvaila}},\ }\bibfield  {title} {\bibinfo {title} {Towards quantum
  technologies with gamma photons},\ }\href
  {https://doi.org/10.1140/epjqt/s40507-024-00240-2} {\bibfield  {journal}
  {\bibinfo  {journal} {EPJ Quantum Technol.}\ }\textbf {\bibinfo {volume}
  {11}},\ \bibinfo {pages} {39} (\bibinfo {year} {2024})}\BibitemShut {NoStop}%
\bibitem [{\citenamefont {Xu}\ \emph {et~al.}(2024)\citenamefont {Xu},
  \citenamefont {Balabanski}, \citenamefont {Baran}, \citenamefont {Iorga},\
  and\ \citenamefont {Matei}}]{Xu:2024jlt}%
  \BibitemOpen
  \bibfield  {author} {\bibinfo {author} {\bibfnamefont {Y.}~\bibnamefont
  {Xu}}, \bibinfo {author} {\bibfnamefont {D.~L.}\ \bibnamefont {Balabanski}},
  \bibinfo {author} {\bibfnamefont {V.}~\bibnamefont {Baran}}, \bibinfo
  {author} {\bibfnamefont {C.}~\bibnamefont {Iorga}},\ and\ \bibinfo {author}
  {\bibfnamefont {C.}~\bibnamefont {Matei}},\ }\bibfield  {title} {\bibinfo
  {title} {Vortex photon induced nuclear reaction: Mechanism, model, and
  application to the studies of giant resonance and astrophysical reaction
  rate},\ }\href {https://doi.org/10.1016/j.physletb.2024.138622} {\bibfield
  {journal} {\bibinfo  {journal} {Phys. Lett. B}\ }\textbf {\bibinfo {volume}
  {852}},\ \bibinfo {pages} {138622} (\bibinfo {year} {2024})}\BibitemShut
  {NoStop}%
\bibitem [{\citenamefont {Klein}\ and\ \citenamefont
  {Mäntysaari}(2019)}]{Klein2019}%
  \BibitemOpen
  \bibfield  {author} {\bibinfo {author} {\bibfnamefont {S.~R.}\ \bibnamefont
  {Klein}}\ and\ \bibinfo {author} {\bibfnamefont {H.}~\bibnamefont
  {Mäntysaari}},\ }\bibfield  {title} {\bibinfo {title} {Imaging the nucleus
  with high-energy photons},\ }\href
  {https://doi.org/10.1038/s42254-019-0107-6} {\bibfield  {journal} {\bibinfo
  {journal} {Nat. Rev. Phys.}\ }\textbf {\bibinfo {volume} {1}},\ \bibinfo
  {pages} {662} (\bibinfo {year} {2019})}\BibitemShut {NoStop}%
\bibitem [{\citenamefont {Ivanov}(2022)}]{Ivanov:2022jzh}%
  \BibitemOpen
  \bibfield  {author} {\bibinfo {author} {\bibfnamefont {I.~P.}\ \bibnamefont
  {Ivanov}},\ }\bibfield  {title} {\bibinfo {title} {Promises and challenges of
  high-energy vortex states collisions},\ }\href
  {https://doi.org/10.1016/j.ppnp.2022.103987} {\bibfield  {journal} {\bibinfo
  {journal} {Prog. Part. Nucl. Phys.}\ }\textbf {\bibinfo {volume} {127}},\
  \bibinfo {pages} {103987} (\bibinfo {year} {2022})}\BibitemShut {NoStop}%
\bibitem [{\citenamefont {Zilges}\ \emph {et~al.}(2022)\citenamefont {Zilges},
  \citenamefont {Balabanski}, \citenamefont {Isaak},\ and\ \citenamefont
  {Pietralla}}]{Zilges:2022ugq}%
  \BibitemOpen
  \bibfield  {author} {\bibinfo {author} {\bibfnamefont {A.}~\bibnamefont
  {Zilges}}, \bibinfo {author} {\bibfnamefont {D.}~\bibnamefont {Balabanski}},
  \bibinfo {author} {\bibfnamefont {J.}~\bibnamefont {Isaak}},\ and\ \bibinfo
  {author} {\bibfnamefont {N.}~\bibnamefont {Pietralla}},\ }\bibfield  {title}
  {\bibinfo {title} {Photonuclear reactions—from basic research to
  applications},\ }\href
  {https://doi.org/https://doi.org/10.1016/j.ppnp.2021.103903} {\bibfield
  {journal} {\bibinfo  {journal} {Prog. Part. Nucl. Phys.}\ }\textbf {\bibinfo
  {volume} {122}},\ \bibinfo {pages} {103903} (\bibinfo {year}
  {2022})}\BibitemShut {NoStop}%
\bibitem [{\citenamefont {Liu}\ and\ \citenamefont {Ji}(2026)}]{5md2-ngcf2026}%
  \BibitemOpen
  \bibfield  {author} {\bibinfo {author} {\bibfnamefont {S.}~\bibnamefont
  {Liu}}\ and\ \bibinfo {author} {\bibfnamefont {L.}~\bibnamefont {Ji}},\
  }\bibfield  {title} {\bibinfo {title} {Vortex-photon-induced multipole
  transitions in atomic nuclei},\ }\href {https://doi.org/10.1103/5md2-ngcf}
  {\bibfield  {journal} {\bibinfo  {journal} {Phys. Rev. A}\ }\textbf {\bibinfo
  {volume} {113}},\ \bibinfo {pages} {013121} (\bibinfo {year}
  {2026})}\BibitemShut {NoStop}%
\bibitem [{\citenamefont {Sherwin}(2017)}]{PhysRevA.96.062120}%
  \BibitemOpen
  \bibfield  {author} {\bibinfo {author} {\bibfnamefont {J.~A.}\ \bibnamefont
  {Sherwin}},\ }\bibfield  {title} {\bibinfo {title} {Compton scattering of
  bessel light with large recoil parameter},\ }\href
  {https://doi.org/10.1103/PhysRevA.96.062120} {\bibfield  {journal} {\bibinfo
  {journal} {Phys. Rev. A}\ }\textbf {\bibinfo {volume} {96}},\ \bibinfo
  {pages} {062120} (\bibinfo {year} {2017})}\BibitemShut {NoStop}%
\bibitem [{\citenamefont {Aboushelbaya}\ \emph {et~al.}(2019)\citenamefont
  {Aboushelbaya}, \citenamefont {Glize}, \citenamefont {Savin}, \citenamefont
  {Mayr}, \citenamefont {Spiers}, \citenamefont {Wang}, \citenamefont
  {Collier}, \citenamefont {Marklund}, \citenamefont {Trines}, \citenamefont
  {Bingham} \emph {et~al.}}]{PhysRevLett.123.113604}%
  \BibitemOpen
  \bibfield  {author} {\bibinfo {author} {\bibfnamefont {R.}~\bibnamefont
  {Aboushelbaya}}, \bibinfo {author} {\bibfnamefont {K.}~\bibnamefont {Glize}},
  \bibinfo {author} {\bibfnamefont {A.~F.}\ \bibnamefont {Savin}}, \bibinfo
  {author} {\bibfnamefont {M.}~\bibnamefont {Mayr}}, \bibinfo {author}
  {\bibfnamefont {B.}~\bibnamefont {Spiers}}, \bibinfo {author} {\bibfnamefont
  {R.}~\bibnamefont {Wang}}, \bibinfo {author} {\bibfnamefont {J.}~\bibnamefont
  {Collier}}, \bibinfo {author} {\bibfnamefont {M.}~\bibnamefont {Marklund}},
  \bibinfo {author} {\bibfnamefont {R.~M. G.~M.}\ \bibnamefont {Trines}},
  \bibinfo {author} {\bibfnamefont {R.}~\bibnamefont {Bingham}}, \emph
  {et~al.},\ }\bibfield  {title} {\bibinfo {title} {Orbital angular momentum
  coupling in elastic photon-photon scattering},\ }\href
  {https://doi.org/10.1103/PhysRevLett.123.113604} {\bibfield  {journal}
  {\bibinfo  {journal} {Phys. Rev. Lett.}\ }\textbf {\bibinfo {volume} {123}},\
  \bibinfo {pages} {113604} (\bibinfo {year} {2019})}\BibitemShut {NoStop}%
\bibitem [{\citenamefont {Bu}\ \emph {et~al.}(2021)\citenamefont {Bu},
  \citenamefont {Ji}, \citenamefont {Lei}, \citenamefont {Hu}, \citenamefont
  {Zhang},\ and\ \citenamefont {Shen}}]{PhysRevResearch.3.043159}%
  \BibitemOpen
  \bibfield  {author} {\bibinfo {author} {\bibfnamefont {Z.}~\bibnamefont
  {Bu}}, \bibinfo {author} {\bibfnamefont {L.}~\bibnamefont {Ji}}, \bibinfo
  {author} {\bibfnamefont {S.}~\bibnamefont {Lei}}, \bibinfo {author}
  {\bibfnamefont {H.}~\bibnamefont {Hu}}, \bibinfo {author} {\bibfnamefont
  {X.}~\bibnamefont {Zhang}},\ and\ \bibinfo {author} {\bibfnamefont
  {B.}~\bibnamefont {Shen}},\ }\bibfield  {title} {\bibinfo {title} {Twisted
  breit-wheeler electron-positron pair creation via vortex gamma photons},\
  }\href {https://doi.org/10.1103/PhysRevResearch.3.043159} {\bibfield
  {journal} {\bibinfo  {journal} {Phys. Rev. Res.}\ }\textbf {\bibinfo {volume}
  {3}},\ \bibinfo {pages} {043159} (\bibinfo {year} {2021})}\BibitemShut
  {NoStop}%
\bibitem [{\citenamefont {Ivanov}\ \emph {et~al.}(2020)\citenamefont {Ivanov},
  \citenamefont {Korchagin}, \citenamefont {Pimikov},\ and\ \citenamefont
  {Zhang}}]{Ivanov:2019vxe}%
  \BibitemOpen
  \bibfield  {author} {\bibinfo {author} {\bibfnamefont {I.~P.}\ \bibnamefont
  {Ivanov}}, \bibinfo {author} {\bibfnamefont {N.}~\bibnamefont {Korchagin}},
  \bibinfo {author} {\bibfnamefont {A.}~\bibnamefont {Pimikov}},\ and\ \bibinfo
  {author} {\bibfnamefont {P.}~\bibnamefont {Zhang}},\ }\bibfield  {title}
  {\bibinfo {title} {Doing spin physics with unpolarized particles},\ }\href
  {https://doi.org/10.1103/PhysRevLett.124.192001} {\bibfield  {journal}
  {\bibinfo  {journal} {Phys. Rev. Lett.}\ }\textbf {\bibinfo {volume} {124}},\
  \bibinfo {pages} {192001} (\bibinfo {year} {2020})}\BibitemShut {NoStop}%
\bibitem [{\citenamefont {Afanasev}\ \emph {et~al.}(2021)\citenamefont
  {Afanasev}, \citenamefont {Carlson},\ and\ \citenamefont
  {Mukherjee}}]{afanasev2021recoil}%
  \BibitemOpen
  \bibfield  {author} {\bibinfo {author} {\bibfnamefont {A.}~\bibnamefont
  {Afanasev}}, \bibinfo {author} {\bibfnamefont {C.~E.}\ \bibnamefont
  {Carlson}},\ and\ \bibinfo {author} {\bibfnamefont {A.}~\bibnamefont
  {Mukherjee}},\ }\bibfield  {title} {\bibinfo {title} {Recoil momentum effects
  in quantum processes induced by twisted photons},\ }\href
  {https://doi.org/10.1103/PhysRevResearch.3.023097} {\bibfield  {journal}
  {\bibinfo  {journal} {Phys. Rev. Res.}\ }\textbf {\bibinfo {volume} {3}},\
  \bibinfo {pages} {023097} (\bibinfo {year} {2021})}\BibitemShut {NoStop}%
\bibitem [{\citenamefont {Surzhykov}\ \emph {et~al.}(2015)\citenamefont
  {Surzhykov}, \citenamefont {Seipt}, \citenamefont {Serbo},\ and\
  \citenamefont {Fritzsche}}]{PhysRevA.91.013403}%
  \BibitemOpen
  \bibfield  {author} {\bibinfo {author} {\bibfnamefont {A.}~\bibnamefont
  {Surzhykov}}, \bibinfo {author} {\bibfnamefont {D.}~\bibnamefont {Seipt}},
  \bibinfo {author} {\bibfnamefont {V.~G.}\ \bibnamefont {Serbo}},\ and\
  \bibinfo {author} {\bibfnamefont {S.}~\bibnamefont {Fritzsche}},\ }\bibfield
  {title} {\bibinfo {title} {Interaction of twisted light with many-electron
  atoms and ions},\ }\href {https://doi.org/10.1103/PhysRevA.91.013403}
  {\bibfield  {journal} {\bibinfo  {journal} {Phys. Rev. A}\ }\textbf {\bibinfo
  {volume} {91}},\ \bibinfo {pages} {013403} (\bibinfo {year}
  {2015})}\BibitemShut {NoStop}%
\bibitem [{\citenamefont {Xiao}\ \emph {et~al.}(2016)\citenamefont {Xiao},
  \citenamefont {Klitis}, \citenamefont {Li}, \citenamefont {Chen},
  \citenamefont {Cai}, \citenamefont {Sorel},\ and\ \citenamefont
  {Yu}}]{xiaoGenerationPhotonicOrbital2016}%
  \BibitemOpen
  \bibfield  {author} {\bibinfo {author} {\bibfnamefont {Q.}~\bibnamefont
  {Xiao}}, \bibinfo {author} {\bibfnamefont {C.}~\bibnamefont {Klitis}},
  \bibinfo {author} {\bibfnamefont {S.}~\bibnamefont {Li}}, \bibinfo {author}
  {\bibfnamefont {Y.}~\bibnamefont {Chen}}, \bibinfo {author} {\bibfnamefont
  {X.}~\bibnamefont {Cai}}, \bibinfo {author} {\bibfnamefont {M.}~\bibnamefont
  {Sorel}},\ and\ \bibinfo {author} {\bibfnamefont {S.}~\bibnamefont {Yu}},\
  }\bibfield  {title} {\bibinfo {title} {Generation of photonic orbital angular
  momentum superposition states using vortex beam emitters with superimposed
  gratings},\ }\href {https://doi.org/10.1364/OE.24.003168} {\bibfield
  {journal} {\bibinfo  {journal} {Opt. Express}\ }\textbf {\bibinfo {volume}
  {24}},\ \bibinfo {pages} {3168} (\bibinfo {year} {2016})}\BibitemShut
  {NoStop}%
\bibitem [{\citenamefont {Knyazev}\ and\ \citenamefont
  {Serbo}(2018)}]{knyazev2018beams}%
  \BibitemOpen
  \bibfield  {author} {\bibinfo {author} {\bibfnamefont {B.~A.}\ \bibnamefont
  {Knyazev}}\ and\ \bibinfo {author} {\bibfnamefont {V.}~\bibnamefont
  {Serbo}},\ }\bibfield  {title} {\bibinfo {title} {Beams of photons with
  nonzero projections of orbital angular momenta: new results},\ }\href
  {https://doi.org/10.3367/UFNe.2018.02.038306} {\bibfield  {journal} {\bibinfo
   {journal} {Phys. Usp.}\ }\textbf {\bibinfo {volume} {61}},\ \bibinfo {pages}
  {449} (\bibinfo {year} {2018})}\BibitemShut {NoStop}%
\bibitem [{\citenamefont {Stock}\ \emph {et~al.}(2015)\citenamefont {Stock},
  \citenamefont {Surzhykov}, \citenamefont {Fritzsche},\ and\ \citenamefont
  {Seipt}}]{PhysRevA.92.013401}%
  \BibitemOpen
  \bibfield  {author} {\bibinfo {author} {\bibfnamefont {S.}~\bibnamefont
  {Stock}}, \bibinfo {author} {\bibfnamefont {A.}~\bibnamefont {Surzhykov}},
  \bibinfo {author} {\bibfnamefont {S.}~\bibnamefont {Fritzsche}},\ and\
  \bibinfo {author} {\bibfnamefont {D.}~\bibnamefont {Seipt}},\ }\bibfield
  {title} {\bibinfo {title} {Compton scattering of twisted light: Angular
  distribution and polarization of scattered photons},\ }\href
  {https://doi.org/10.1103/PhysRevA.92.013401} {\bibfield  {journal} {\bibinfo
  {journal} {Phys. Rev. A}\ }\textbf {\bibinfo {volume} {92}},\ \bibinfo
  {pages} {013401} (\bibinfo {year} {2015})}\BibitemShut {NoStop}%
\bibitem [{\citenamefont {Ababekri}\ \emph
  {et~al.}(2024{\natexlab{a}})\citenamefont {Ababekri}, \citenamefont {Zhou},
  \citenamefont {Guo}, \citenamefont {Ren}, \citenamefont {Kou}, \citenamefont
  {Zhao}, \citenamefont {Li},\ and\ \citenamefont {Li}}]{Ababekri:2024cyd}%
  \BibitemOpen
  \bibfield  {author} {\bibinfo {author} {\bibfnamefont {M.}~\bibnamefont
  {Ababekri}}, \bibinfo {author} {\bibfnamefont {J.-L.}\ \bibnamefont {Zhou}},
  \bibinfo {author} {\bibfnamefont {R.-T.}\ \bibnamefont {Guo}}, \bibinfo
  {author} {\bibfnamefont {Y.-Z.}\ \bibnamefont {Ren}}, \bibinfo {author}
  {\bibfnamefont {Y.-H.}\ \bibnamefont {Kou}}, \bibinfo {author} {\bibfnamefont
  {Q.}~\bibnamefont {Zhao}}, \bibinfo {author} {\bibfnamefont {Z.-P.}\
  \bibnamefont {Li}},\ and\ \bibinfo {author} {\bibfnamefont {J.-X.}\
  \bibnamefont {Li}},\ }\bibfield  {title} {\bibinfo {title} {Generation of
  ultrarelativistic vortex leptons with large orbital angular momenta},\ }\href
  {https://doi.org/10.1103/PhysRevD.110.076024} {\bibfield  {journal} {\bibinfo
   {journal} {Phys. Rev. D}\ }\textbf {\bibinfo {volume} {110}},\ \bibinfo
  {pages} {076024} (\bibinfo {year} {2024}{\natexlab{a}})}\BibitemShut
  {NoStop}%
\bibitem [{\citenamefont {Vasilyeu}\ \emph {et~al.}(2009)\citenamefont
  {Vasilyeu}, \citenamefont {Dudley}, \citenamefont {Khilo},\ and\
  \citenamefont {Forbes}}]{vasilyeu_generating_2009}%
  \BibitemOpen
  \bibfield  {author} {\bibinfo {author} {\bibfnamefont {R.}~\bibnamefont
  {Vasilyeu}}, \bibinfo {author} {\bibfnamefont {A.}~\bibnamefont {Dudley}},
  \bibinfo {author} {\bibfnamefont {N.}~\bibnamefont {Khilo}},\ and\ \bibinfo
  {author} {\bibfnamefont {A.}~\bibnamefont {Forbes}},\ }\bibfield  {title}
  {\bibinfo {title} {Generating superpositions of higher-order bessel beams},\
  }\href {https://doi.org/10.1364/OE.17.023389} {\bibfield  {journal} {\bibinfo
   {journal} {Opt. Express}\ }\textbf {\bibinfo {volume} {17}},\ \bibinfo
  {pages} {23389} (\bibinfo {year} {2009})}\BibitemShut {NoStop}%
\bibitem [{\citenamefont {Li}\ \emph {et~al.}(2015)\citenamefont {Li},
  \citenamefont {Strain}, \citenamefont {Meriggi}, \citenamefont {Chen},
  \citenamefont {Zhu}, \citenamefont {Cicek}, \citenamefont {Wang},
  \citenamefont {Cai}, \citenamefont {Sorel}, \citenamefont {Thompson} \emph
  {et~al.}}]{liPatternManipulationOnchip2015}%
  \BibitemOpen
  \bibfield  {author} {\bibinfo {author} {\bibfnamefont {H.}~\bibnamefont
  {Li}}, \bibinfo {author} {\bibfnamefont {M.~J.}\ \bibnamefont {Strain}},
  \bibinfo {author} {\bibfnamefont {L.}~\bibnamefont {Meriggi}}, \bibinfo
  {author} {\bibfnamefont {L.}~\bibnamefont {Chen}}, \bibinfo {author}
  {\bibfnamefont {J.}~\bibnamefont {Zhu}}, \bibinfo {author} {\bibfnamefont
  {K.}~\bibnamefont {Cicek}}, \bibinfo {author} {\bibfnamefont
  {J.}~\bibnamefont {Wang}}, \bibinfo {author} {\bibfnamefont {X.}~\bibnamefont
  {Cai}}, \bibinfo {author} {\bibfnamefont {M.}~\bibnamefont {Sorel}}, \bibinfo
  {author} {\bibfnamefont {M.~G.}\ \bibnamefont {Thompson}}, \emph {et~al.},\
  }\bibfield  {title} {\bibinfo {title} {Pattern manipulation via on-chip phase
  modulation between orbital angular momentum beams},\ }\href
  {https://doi.org/10.1063/1.4927758} {\bibfield  {journal} {\bibinfo
  {journal} {Appl. Phys. Lett.}\ }\textbf {\bibinfo {volume} {107}},\ \bibinfo
  {pages} {051102} (\bibinfo {year} {2015})}\BibitemShut {NoStop}%
\bibitem [{\citenamefont {Rubinsztein-Dunlop}\ \emph
  {et~al.}(2017)\citenamefont {Rubinsztein-Dunlop}, \citenamefont {Forbes},
  \citenamefont {Berry}, \citenamefont {Dennis}, \citenamefont {Andrews},
  \citenamefont {Mansuripur}, \citenamefont {Denz}, \citenamefont {Alpmann},
  \citenamefont {Banzer}, \citenamefont {Bauer} \emph
  {et~al.}}]{RubinszteinDunlop2017}%
  \BibitemOpen
  \bibfield  {author} {\bibinfo {author} {\bibfnamefont {H.}~\bibnamefont
  {Rubinsztein-Dunlop}}, \bibinfo {author} {\bibfnamefont {A.}~\bibnamefont
  {Forbes}}, \bibinfo {author} {\bibfnamefont {M.~V.}\ \bibnamefont {Berry}},
  \bibinfo {author} {\bibfnamefont {M.~R.}\ \bibnamefont {Dennis}}, \bibinfo
  {author} {\bibfnamefont {D.~L.}\ \bibnamefont {Andrews}}, \bibinfo {author}
  {\bibfnamefont {M.}~\bibnamefont {Mansuripur}}, \bibinfo {author}
  {\bibfnamefont {C.}~\bibnamefont {Denz}}, \bibinfo {author} {\bibfnamefont
  {C.}~\bibnamefont {Alpmann}}, \bibinfo {author} {\bibfnamefont
  {P.}~\bibnamefont {Banzer}}, \bibinfo {author} {\bibfnamefont
  {T.}~\bibnamefont {Bauer}}, \emph {et~al.},\ }\bibfield  {title} {\bibinfo
  {title} {Roadmap on structured light},\ }\href
  {https://doi.org/10.1088/2040-8978/19/1/013001} {\bibfield  {journal}
  {\bibinfo  {journal} {J. Opt.}\ }\textbf {\bibinfo {volume} {19}},\ \bibinfo
  {pages} {013001} (\bibinfo {year} {2017})}\BibitemShut {NoStop}%
\bibitem [{\citenamefont {Lee}\ \emph {et~al.}(2019)\citenamefont {Lee},
  \citenamefont {Alexander}, \citenamefont {Kevan}, \citenamefont {Roy},\ and\
  \citenamefont {McMorran}}]{lee_laguerregauss_2019}%
  \BibitemOpen
  \bibfield  {author} {\bibinfo {author} {\bibfnamefont {J.~C.~T.}\
  \bibnamefont {Lee}}, \bibinfo {author} {\bibfnamefont {S.~J.}\ \bibnamefont
  {Alexander}}, \bibinfo {author} {\bibfnamefont {S.~D.}\ \bibnamefont
  {Kevan}}, \bibinfo {author} {\bibfnamefont {S.}~\bibnamefont {Roy}},\ and\
  \bibinfo {author} {\bibfnamefont {B.~J.}\ \bibnamefont {McMorran}},\
  }\bibfield  {title} {\bibinfo {title} {Laguerre–gauss and hermite–gauss
  soft x-ray states generated using diffractive optics},\ }\href
  {https://doi.org/10.1038/s41566-018-0328-8} {\bibfield  {journal} {\bibinfo
  {journal} {Nat. Photonics}\ }\textbf {\bibinfo {volume} {13}},\ \bibinfo
  {pages} {205} (\bibinfo {year} {2019})}\BibitemShut {NoStop}%
\bibitem [{\citenamefont {Zheng}\ \emph {et~al.}(2021)\citenamefont {Zheng},
  \citenamefont {Wang}, \citenamefont {Li}, \citenamefont {Li}, \citenamefont
  {Wang}, \citenamefont {Zhao}, \citenamefont {Li}, \citenamefont {Yue},
  \citenamefont {Zhang}, \citenamefont {Zhang} \emph
  {et~al.}}]{yao_alldielectric_2021}%
  \BibitemOpen
  \bibfield  {author} {\bibinfo {author} {\bibfnamefont {C.}~\bibnamefont
  {Zheng}}, \bibinfo {author} {\bibfnamefont {G.}~\bibnamefont {Wang}},
  \bibinfo {author} {\bibfnamefont {J.}~\bibnamefont {Li}}, \bibinfo {author}
  {\bibfnamefont {J.}~\bibnamefont {Li}}, \bibinfo {author} {\bibfnamefont
  {S.}~\bibnamefont {Wang}}, \bibinfo {author} {\bibfnamefont {H.}~\bibnamefont
  {Zhao}}, \bibinfo {author} {\bibfnamefont {M.}~\bibnamefont {Li}}, \bibinfo
  {author} {\bibfnamefont {Z.}~\bibnamefont {Yue}}, \bibinfo {author}
  {\bibfnamefont {Y.}~\bibnamefont {Zhang}}, \bibinfo {author} {\bibfnamefont
  {Y.}~\bibnamefont {Zhang}}, \emph {et~al.},\ }\bibfield  {title} {\bibinfo
  {title} {All-dielectric metasurface for manipulating the superpositions of
  orbital angular momentum via spin-decoupling},\ }\href
  {https://doi.org/10.1002/adom.202002007} {\bibfield  {journal} {\bibinfo
  {journal} {Adv. Opt. Mater.}\ }\textbf {\bibinfo {volume} {9}},\ \bibinfo
  {pages} {2002007} (\bibinfo {year} {2021})}\BibitemShut {NoStop}%
\bibitem [{\citenamefont {Liu}\ \emph {et~al.}(2024)\citenamefont {Liu},
  \citenamefont {Yan}, \citenamefont {Afanasev}, \citenamefont {Benson},
  \citenamefont {Hao}, \citenamefont {Mikhailov}, \citenamefont {Popov},\ and\
  \citenamefont {Wu}}]{liu_generation_2024}%
  \BibitemOpen
  \bibfield  {author} {\bibinfo {author} {\bibfnamefont {P.}~\bibnamefont
  {Liu}}, \bibinfo {author} {\bibfnamefont {J.}~\bibnamefont {Yan}}, \bibinfo
  {author} {\bibfnamefont {A.}~\bibnamefont {Afanasev}}, \bibinfo {author}
  {\bibfnamefont {S.~V.}\ \bibnamefont {Benson}}, \bibinfo {author}
  {\bibfnamefont {H.}~\bibnamefont {Hao}}, \bibinfo {author} {\bibfnamefont
  {S.~F.}\ \bibnamefont {Mikhailov}}, \bibinfo {author} {\bibfnamefont {V.~G.}\
  \bibnamefont {Popov}},\ and\ \bibinfo {author} {\bibfnamefont {Y.~K.}\
  \bibnamefont {Wu}},\ }\bibfield  {title} {\bibinfo {title} {Generation of
  superposed orbital angular momentum beams using a free-electron laser
  oscillator},\ }\href {https://doi.org/10.1364/OE.510649} {\bibfield
  {journal} {\bibinfo  {journal} {Opt. Express}\ }\textbf {\bibinfo {volume}
  {32}},\ \bibinfo {pages} {2235} (\bibinfo {year} {2024})}\BibitemShut
  {NoStop}%
\bibitem [{\citenamefont {Jentschura}\ and\ \citenamefont
  {Serbo}(2011{\natexlab{a}})}]{Jentschura:2010ap}%
  \BibitemOpen
  \bibfield  {author} {\bibinfo {author} {\bibfnamefont {U.~D.}\ \bibnamefont
  {Jentschura}}\ and\ \bibinfo {author} {\bibfnamefont {V.~G.}\ \bibnamefont
  {Serbo}},\ }\bibfield  {title} {\bibinfo {title} {Generation of high-energy
  photons with large orbital angular momentum by compton backscattering},\
  }\href {https://doi.org/10.1103/PhysRevLett.106.013001} {\bibfield  {journal}
  {\bibinfo  {journal} {Phys. Rev. Lett.}\ }\textbf {\bibinfo {volume} {106}},\
  \bibinfo {pages} {013001} (\bibinfo {year} {2011}{\natexlab{a}})}\BibitemShut
  {NoStop}%
\bibitem [{\citenamefont {Petrillo}\ \emph {et~al.}(2016)\citenamefont
  {Petrillo}, \citenamefont {Dattoli}, \citenamefont {Drebot},\ and\
  \citenamefont {Nguyen}}]{petrilloComptonScatteredXGamma2016}%
  \BibitemOpen
  \bibfield  {author} {\bibinfo {author} {\bibfnamefont {V.}~\bibnamefont
  {Petrillo}}, \bibinfo {author} {\bibfnamefont {G.}~\bibnamefont {Dattoli}},
  \bibinfo {author} {\bibfnamefont {I.}~\bibnamefont {Drebot}},\ and\ \bibinfo
  {author} {\bibfnamefont {F.}~\bibnamefont {Nguyen}},\ }\bibfield  {title}
  {\bibinfo {title} {Compton scattered x-gamma rays with orbital momentum},\
  }\href {https://doi.org/10.1103/PhysRevLett.117.123903} {\bibfield  {journal}
  {\bibinfo  {journal} {Phys. Rev. Lett.}\ }\textbf {\bibinfo {volume} {117}},\
  \bibinfo {pages} {123903} (\bibinfo {year} {2016})}\BibitemShut {NoStop}%
\bibitem [{\citenamefont {Taira}\ \emph {et~al.}(2017)\citenamefont {Taira},
  \citenamefont {Hayakawa},\ and\ \citenamefont {Katoh}}]{taira2017gamma}%
  \BibitemOpen
  \bibfield  {author} {\bibinfo {author} {\bibfnamefont {Y.}~\bibnamefont
  {Taira}}, \bibinfo {author} {\bibfnamefont {T.}~\bibnamefont {Hayakawa}},\
  and\ \bibinfo {author} {\bibfnamefont {M.}~\bibnamefont {Katoh}},\ }\bibfield
   {title} {\bibinfo {title} {Gamma-ray vortices from nonlinear inverse thomson
  scattering of circularly polarized light},\ }\href
  {https://doi.org/10.1038/s41598-017-05187-2} {\bibfield  {journal} {\bibinfo
  {journal} {Sci. Rep.}\ }\textbf {\bibinfo {volume} {7}},\ \bibinfo {pages}
  {5018} (\bibinfo {year} {2017})}\BibitemShut {NoStop}%
\bibitem [{\citenamefont {Ababekri}\ \emph
  {et~al.}(2024{\natexlab{b}})\citenamefont {Ababekri}, \citenamefont {Guo},
  \citenamefont {Wan}, \citenamefont {Qiao}, \citenamefont {Li}, \citenamefont
  {Lv}, \citenamefont {Zhang}, \citenamefont {Zhou}, \citenamefont {Gu},\ and\
  \citenamefont {Li}}]{Ababekri:2022mob}%
  \BibitemOpen
  \bibfield  {author} {\bibinfo {author} {\bibfnamefont {M.}~\bibnamefont
  {Ababekri}}, \bibinfo {author} {\bibfnamefont {R.-T.}\ \bibnamefont {Guo}},
  \bibinfo {author} {\bibfnamefont {F.}~\bibnamefont {Wan}}, \bibinfo {author}
  {\bibfnamefont {B.}~\bibnamefont {Qiao}}, \bibinfo {author} {\bibfnamefont
  {Z.}~\bibnamefont {Li}}, \bibinfo {author} {\bibfnamefont {C.}~\bibnamefont
  {Lv}}, \bibinfo {author} {\bibfnamefont {B.}~\bibnamefont {Zhang}}, \bibinfo
  {author} {\bibfnamefont {W.}~\bibnamefont {Zhou}}, \bibinfo {author}
  {\bibfnamefont {Y.}~\bibnamefont {Gu}},\ and\ \bibinfo {author}
  {\bibfnamefont {J.-X.}\ \bibnamefont {Li}},\ }\bibfield  {title} {\bibinfo
  {title} {Vortex $\ensuremath{\gamma}$ photon generation via spin-to-orbital
  angular momentum transfer in nonlinear compton scattering},\ }\href
  {https://doi.org/10.1103/PhysRevD.109.016005} {\bibfield  {journal} {\bibinfo
   {journal} {Phys. Rev. D}\ }\textbf {\bibinfo {volume} {109}},\ \bibinfo
  {pages} {016005} (\bibinfo {year} {2024}{\natexlab{b}})}\BibitemShut
  {NoStop}%
\bibitem [{\citenamefont {Jiang}\ \emph {et~al.}(2025)\citenamefont {Jiang},
  \citenamefont {Zhuang}, \citenamefont {Chen}, \citenamefont {Li},\ and\
  \citenamefont {Chen}}]{PhysRevLett.134.153802}%
  \BibitemOpen
  \bibfield  {author} {\bibinfo {author} {\bibfnamefont {J.-J.}\ \bibnamefont
  {Jiang}}, \bibinfo {author} {\bibfnamefont {K.-H.}\ \bibnamefont {Zhuang}},
  \bibinfo {author} {\bibfnamefont {J.-D.}\ \bibnamefont {Chen}}, \bibinfo
  {author} {\bibfnamefont {J.-X.}\ \bibnamefont {Li}},\ and\ \bibinfo {author}
  {\bibfnamefont {Y.-Y.}\ \bibnamefont {Chen}},\ }\bibfield  {title} {\bibinfo
  {title} {Controlling the polarization and vortex charge of
  $\ensuremath{\gamma}$ photons via nonlinear compton scattering},\ }\href
  {https://doi.org/10.1103/PhysRevLett.134.153802} {\bibfield  {journal}
  {\bibinfo  {journal} {Phys. Rev. Lett.}\ }\textbf {\bibinfo {volume} {134}},\
  \bibinfo {pages} {153802} (\bibinfo {year} {2025})}\BibitemShut {NoStop}%
\bibitem [{\citenamefont {Chen}\ \emph {et~al.}(2018)\citenamefont {Chen},
  \citenamefont {Li}, \citenamefont {Hatsagortsyan},\ and\ \citenamefont
  {Keitel}}]{Chen:2018tkb}%
  \BibitemOpen
  \bibfield  {author} {\bibinfo {author} {\bibfnamefont {Y.-Y.}\ \bibnamefont
  {Chen}}, \bibinfo {author} {\bibfnamefont {J.-X.}\ \bibnamefont {Li}},
  \bibinfo {author} {\bibfnamefont {K.~Z.}\ \bibnamefont {Hatsagortsyan}},\
  and\ \bibinfo {author} {\bibfnamefont {C.~H.}\ \bibnamefont {Keitel}},\
  }\bibfield  {title} {\bibinfo {title} {$\ensuremath{\gamma}$-ray beams with
  large orbital angular momentum via nonlinear compton scattering with
  radiation reaction},\ }\href {https://doi.org/10.1103/PhysRevLett.121.074801}
  {\bibfield  {journal} {\bibinfo  {journal} {Phys. Rev. Lett.}\ }\textbf
  {\bibinfo {volume} {121}},\ \bibinfo {pages} {074801} (\bibinfo {year}
  {2018})}\BibitemShut {NoStop}%
\bibitem [{\citenamefont {Liu}\ \emph {et~al.}(2020)\citenamefont {Liu},
  \citenamefont {Salamin}, \citenamefont {Dou}, \citenamefont {Xu},\ and\
  \citenamefont {Li}}]{liuVortexRaysScattering2020}%
  \BibitemOpen
  \bibfield  {author} {\bibinfo {author} {\bibfnamefont {Y.-Y.}\ \bibnamefont
  {Liu}}, \bibinfo {author} {\bibfnamefont {Y.~I.}\ \bibnamefont {Salamin}},
  \bibinfo {author} {\bibfnamefont {Z.-K.}\ \bibnamefont {Dou}}, \bibinfo
  {author} {\bibfnamefont {Z.-F.}\ \bibnamefont {Xu}},\ and\ \bibinfo {author}
  {\bibfnamefont {J.-X.}\ \bibnamefont {Li}},\ }\bibfield  {title} {\bibinfo
  {title} {Vortex $\ensuremath{\gamma}$ rays from scattering laser bullets off
  ultrarelativistic electrons},\ }\href {https://doi.org/10.1364/OL.45.000395}
  {\bibfield  {journal} {\bibinfo  {journal} {Opt. Lett.}\ }\textbf {\bibinfo
  {volume} {45}},\ \bibinfo {pages} {395} (\bibinfo {year} {2020})}\BibitemShut
  {NoStop}%
\bibitem [{\citenamefont {Wang}\ \emph {et~al.}(2020)\citenamefont {Wang},
  \citenamefont {Li}, \citenamefont {Gan}, \citenamefont {Xie}, \citenamefont
  {Zhong}, \citenamefont {Zhou}, \citenamefont {Zhu}, \citenamefont {He},\ and\
  \citenamefont {Qiao}}]{wangGenerationIntenseVortex2020}%
  \BibitemOpen
  \bibfield  {author} {\bibinfo {author} {\bibfnamefont {J.}~\bibnamefont
  {Wang}}, \bibinfo {author} {\bibfnamefont {X.}~\bibnamefont {Li}}, \bibinfo
  {author} {\bibfnamefont {L.}~\bibnamefont {Gan}}, \bibinfo {author}
  {\bibfnamefont {Y.}~\bibnamefont {Xie}}, \bibinfo {author} {\bibfnamefont
  {C.}~\bibnamefont {Zhong}}, \bibinfo {author} {\bibfnamefont
  {C.}~\bibnamefont {Zhou}}, \bibinfo {author} {\bibfnamefont {S.}~\bibnamefont
  {Zhu}}, \bibinfo {author} {\bibfnamefont {X.}~\bibnamefont {He}},\ and\
  \bibinfo {author} {\bibfnamefont {B.}~\bibnamefont {Qiao}},\ }\bibfield
  {title} {\bibinfo {title} {Generation of intense vortex gamma rays via
  spin-to-orbital conversion of angular momentum in relativistic laser-plasma
  interactions},\ }\href {https://doi.org/10.1103/PhysRevApplied.14.014094}
  {\bibfield  {journal} {\bibinfo  {journal} {Phys. Rev. Appl.}\ }\textbf
  {\bibinfo {volume} {14}},\ \bibinfo {pages} {014094} (\bibinfo {year}
  {2020})}\BibitemShut {NoStop}%
\bibitem [{\citenamefont {Hu}\ \emph {et~al.}(2021)\citenamefont {Hu},
  \citenamefont {Zhao}, \citenamefont {Zhang}, \citenamefont {Lu},
  \citenamefont {Wang}, \citenamefont {Hu}, \citenamefont {Shao},\ and\
  \citenamefont {Yu}}]{huAttosecondRayVortex2021}%
  \BibitemOpen
  \bibfield  {author} {\bibinfo {author} {\bibfnamefont {Y.-T.}\ \bibnamefont
  {Hu}}, \bibinfo {author} {\bibfnamefont {J.}~\bibnamefont {Zhao}}, \bibinfo
  {author} {\bibfnamefont {H.}~\bibnamefont {Zhang}}, \bibinfo {author}
  {\bibfnamefont {Y.}~\bibnamefont {Lu}}, \bibinfo {author} {\bibfnamefont
  {W.-Q.}\ \bibnamefont {Wang}}, \bibinfo {author} {\bibfnamefont {L.-X.}\
  \bibnamefont {Hu}}, \bibinfo {author} {\bibfnamefont {F.-Q.}\ \bibnamefont
  {Shao}},\ and\ \bibinfo {author} {\bibfnamefont {T.-P.}\ \bibnamefont {Yu}},\
  }\bibfield  {title} {\bibinfo {title} {Attosecond $\ensuremath{\gamma}$-ray
  vortex generation in near-critical-density plasma driven by twisted laser
  pulses},\ }\href {https://doi.org/10.1063/5.0028203} {\bibfield  {journal}
  {\bibinfo  {journal} {Appl. Phys. Lett.}\ }\textbf {\bibinfo {volume}
  {118}},\ \bibinfo {pages} {054101} (\bibinfo {year} {2021})}\BibitemShut
  {NoStop}%
\bibitem [{\citenamefont {Zhang}\ \emph {et~al.}(2021)\citenamefont {Zhang},
  \citenamefont {Zhao}, \citenamefont {Hu}, \citenamefont {Li}, \citenamefont
  {Lu}, \citenamefont {Cao}, \citenamefont {Zou}, \citenamefont {Sheng},
  \citenamefont {Pegoraro}, \citenamefont {McKenna} \emph
  {et~al.}}]{zhangEfficientBrightGray2021}%
  \BibitemOpen
  \bibfield  {author} {\bibinfo {author} {\bibfnamefont {H.}~\bibnamefont
  {Zhang}}, \bibinfo {author} {\bibfnamefont {J.}~\bibnamefont {Zhao}},
  \bibinfo {author} {\bibfnamefont {Y.}~\bibnamefont {Hu}}, \bibinfo {author}
  {\bibfnamefont {Q.}~\bibnamefont {Li}}, \bibinfo {author} {\bibfnamefont
  {Y.}~\bibnamefont {Lu}}, \bibinfo {author} {\bibfnamefont {Y.}~\bibnamefont
  {Cao}}, \bibinfo {author} {\bibfnamefont {D.}~\bibnamefont {Zou}}, \bibinfo
  {author} {\bibfnamefont {Z.}~\bibnamefont {Sheng}}, \bibinfo {author}
  {\bibfnamefont {F.}~\bibnamefont {Pegoraro}}, \bibinfo {author}
  {\bibfnamefont {P.}~\bibnamefont {McKenna}}, \emph {et~al.},\ }\bibfield
  {title} {\bibinfo {title} {Efficient bright ${\gamma}$-ray vortex emission
  from a laser-illuminated light-fan-in-channel target},\ }\href
  {https://doi.org/10.1017/hpl.2021.29} {\bibfield  {journal} {\bibinfo
  {journal} {High Power Laser Sci. Eng.}\ }\textbf {\bibinfo {volume} {9}},\
  \bibinfo {pages} {e43} (\bibinfo {year} {2021})}\BibitemShut {NoStop}%
\bibitem [{\citenamefont {Bake}\ \emph {et~al.}(2022)\citenamefont {Bake},
  \citenamefont {Tang},\ and\ \citenamefont {Xie}}]{bakeBrightGraySource2022}%
  \BibitemOpen
  \bibfield  {author} {\bibinfo {author} {\bibfnamefont {M.~A.}\ \bibnamefont
  {Bake}}, \bibinfo {author} {\bibfnamefont {S.}~\bibnamefont {Tang}},\ and\
  \bibinfo {author} {\bibfnamefont {B.}~\bibnamefont {Xie}},\ }\bibfield
  {title} {\bibinfo {title} {Bright $\ensuremath{\gamma}$-ray source with large
  orbital angular momentum from the laser near-critical-plasma interaction},\
  }\href {https://doi.org/10.1088/2058-6272/ac67bd} {\bibfield  {journal}
  {\bibinfo  {journal} {Plasma Sci. Technol.}\ }\textbf {\bibinfo {volume}
  {24}},\ \bibinfo {pages} {095001} (\bibinfo {year} {2022})}\BibitemShut
  {NoStop}%
\bibitem [{\citenamefont {Bu}\ \emph {et~al.}(2024)\citenamefont {Bu},
  \citenamefont {Ji}, \citenamefont {Geng}, \citenamefont {Liu}, \citenamefont
  {Lei}, \citenamefont {Shen}, \citenamefont {Li},\ and\ \citenamefont
  {Xu}}]{Bu2024}%
  \BibitemOpen
  \bibfield  {author} {\bibinfo {author} {\bibfnamefont {Z.}~\bibnamefont
  {Bu}}, \bibinfo {author} {\bibfnamefont {L.}~\bibnamefont {Ji}}, \bibinfo
  {author} {\bibfnamefont {X.}~\bibnamefont {Geng}}, \bibinfo {author}
  {\bibfnamefont {S.}~\bibnamefont {Liu}}, \bibinfo {author} {\bibfnamefont
  {S.}~\bibnamefont {Lei}}, \bibinfo {author} {\bibfnamefont {B.}~\bibnamefont
  {Shen}}, \bibinfo {author} {\bibfnamefont {R.}~\bibnamefont {Li}},\ and\
  \bibinfo {author} {\bibfnamefont {Z.}~\bibnamefont {Xu}},\ }\bibfield
  {title} {\bibinfo {title} {Generation of quantum vortex electrons with
  intense laser pulses},\ }\href {https://doi.org/10.1002/advs.202404564}
  {\bibfield  {journal} {\bibinfo  {journal} {Adv. Sci.}\ }\textbf {\bibinfo
  {volume} {11}},\ \bibinfo {pages} {2404564} (\bibinfo {year}
  {2024})}\BibitemShut {NoStop}%
\bibitem [{\citenamefont {Wei}\ \emph {et~al.}(2026)\citenamefont {Wei},
  \citenamefont {Chen}, \citenamefont {Wang}, \citenamefont {He}, \citenamefont
  {Hu}, \citenamefont {Zhu}, \citenamefont {Xu}, \citenamefont {Zhou},
  \citenamefont {Jia}, \citenamefont {Ge} \emph {et~al.}}]{Wei:2025zsv}%
  \BibitemOpen
  \bibfield  {author} {\bibinfo {author} {\bibfnamefont {M.}~\bibnamefont
  {Wei}}, \bibinfo {author} {\bibfnamefont {S.}~\bibnamefont {Chen}}, \bibinfo
  {author} {\bibfnamefont {Y.}~\bibnamefont {Wang}}, \bibinfo {author}
  {\bibfnamefont {P.-L.}\ \bibnamefont {He}}, \bibinfo {author} {\bibfnamefont
  {X.}~\bibnamefont {Hu}}, \bibinfo {author} {\bibfnamefont {M.}~\bibnamefont
  {Zhu}}, \bibinfo {author} {\bibfnamefont {H.}~\bibnamefont {Xu}}, \bibinfo
  {author} {\bibfnamefont {W.}~\bibnamefont {Zhou}}, \bibinfo {author}
  {\bibfnamefont {J.}~\bibnamefont {Jia}}, \bibinfo {author} {\bibfnamefont
  {X.}~\bibnamefont {Ge}}, \emph {et~al.},\ }\bibfield  {title} {\bibinfo
  {title} {Experimental evidence of vortex $\gamma$ photons in all-optical
  inverse compton scattering},\ }\href
  {https://doi.org/10.1103/PhysRevLett.136.025001} {\bibfield  {journal}
  {\bibinfo  {journal} {Phys. Rev. Lett.}\ }\textbf {\bibinfo {volume} {136}},\
  \bibinfo {pages} {025001} (\bibinfo {year} {2026})}\BibitemShut {NoStop}%
\bibitem [{\citenamefont {Liu}\ \emph {et~al.}(2026)\citenamefont {Liu},
  \citenamefont {Cao}, \citenamefont {Xue}, \citenamefont {Hu}, \citenamefont
  {Liu}, \citenamefont {Li}, \citenamefont {Li}, \citenamefont {Xu},
  \citenamefont {Liu}, \citenamefont {Wang} \emph {et~al.}}]{Liual._2026}%
  \BibitemOpen
  \bibfield  {author} {\bibinfo {author} {\bibfnamefont {S.-M.}\ \bibnamefont
  {Liu}}, \bibinfo {author} {\bibfnamefont {Y.}~\bibnamefont {Cao}}, \bibinfo
  {author} {\bibfnamefont {K.}~\bibnamefont {Xue}}, \bibinfo {author}
  {\bibfnamefont {L.-X.}\ \bibnamefont {Hu}}, \bibinfo {author} {\bibfnamefont
  {X.-Y.}\ \bibnamefont {Liu}}, \bibinfo {author} {\bibfnamefont {X.-Y.}\
  \bibnamefont {Li}}, \bibinfo {author} {\bibfnamefont {C.-Z.}\ \bibnamefont
  {Li}}, \bibinfo {author} {\bibfnamefont {X.-R.}\ \bibnamefont {Xu}}, \bibinfo
  {author} {\bibfnamefont {K.}~\bibnamefont {Liu}}, \bibinfo {author}
  {\bibfnamefont {W.-Q.}\ \bibnamefont {Wang}}, \emph {et~al.},\ }\bibfield
  {title} {\bibinfo {title} {Generation of polarization-tunable hybrid
  cylindrical vector $\ensuremath{\gamma}$ rays from rotating electron beams},\
  }\href {https://doi.org/10.1017/hpl.2026.10121} {\bibfield  {journal}
  {\bibinfo  {journal} {High Power Laser Sci. Eng.}\ }\textbf {\bibinfo
  {volume} {14}},\ \bibinfo {pages} {e14} (\bibinfo {year} {2026})}\BibitemShut
  {NoStop}%
\bibitem [{\citenamefont {Bahrdt}\ \emph {et~al.}(2013)\citenamefont {Bahrdt},
  \citenamefont {Holldack}, \citenamefont {Kuske}, \citenamefont {M\"uller},
  \citenamefont {Scheer},\ and\ \citenamefont {Schmid}}]{bahrdt2013first}%
  \BibitemOpen
  \bibfield  {author} {\bibinfo {author} {\bibfnamefont {J.}~\bibnamefont
  {Bahrdt}}, \bibinfo {author} {\bibfnamefont {K.}~\bibnamefont {Holldack}},
  \bibinfo {author} {\bibfnamefont {P.}~\bibnamefont {Kuske}}, \bibinfo
  {author} {\bibfnamefont {R.}~\bibnamefont {M\"uller}}, \bibinfo {author}
  {\bibfnamefont {M.}~\bibnamefont {Scheer}},\ and\ \bibinfo {author}
  {\bibfnamefont {P.}~\bibnamefont {Schmid}},\ }\bibfield  {title} {\bibinfo
  {title} {First observation of photons carrying orbital angular momentum in
  undulator radiation},\ }\href
  {https://doi.org/10.1103/PhysRevLett.111.034801} {\bibfield  {journal}
  {\bibinfo  {journal} {Phys. Rev. Lett.}\ }\textbf {\bibinfo {volume} {111}},\
  \bibinfo {pages} {034801} (\bibinfo {year} {2013})}\BibitemShut {NoStop}%
\bibitem [{\citenamefont {Katoh}\ \emph {et~al.}(2017)\citenamefont {Katoh},
  \citenamefont {Fujimoto}, \citenamefont {Kawaguchi}, \citenamefont
  {Tsuchiya}, \citenamefont {Ohmi}, \citenamefont {Kaneyasu}, \citenamefont
  {Taira}, \citenamefont {Hosaka}, \citenamefont {Mochihashi},\ and\
  \citenamefont {Takashima}}]{katoh2017angular}%
  \BibitemOpen
  \bibfield  {author} {\bibinfo {author} {\bibfnamefont {M.}~\bibnamefont
  {Katoh}}, \bibinfo {author} {\bibfnamefont {M.}~\bibnamefont {Fujimoto}},
  \bibinfo {author} {\bibfnamefont {H.}~\bibnamefont {Kawaguchi}}, \bibinfo
  {author} {\bibfnamefont {K.}~\bibnamefont {Tsuchiya}}, \bibinfo {author}
  {\bibfnamefont {K.}~\bibnamefont {Ohmi}}, \bibinfo {author} {\bibfnamefont
  {T.}~\bibnamefont {Kaneyasu}}, \bibinfo {author} {\bibfnamefont
  {Y.}~\bibnamefont {Taira}}, \bibinfo {author} {\bibfnamefont
  {M.}~\bibnamefont {Hosaka}}, \bibinfo {author} {\bibfnamefont
  {A.}~\bibnamefont {Mochihashi}},\ and\ \bibinfo {author} {\bibfnamefont
  {Y.}~\bibnamefont {Takashima}},\ }\bibfield  {title} {\bibinfo {title}
  {Angular momentum of twisted radiation from an electron in spiral motion},\
  }\href {https://doi.org/10.1103/PhysRevLett.118.094801} {\bibfield  {journal}
  {\bibinfo  {journal} {Phys. Rev. Lett.}\ }\textbf {\bibinfo {volume} {118}},\
  \bibinfo {pages} {094801} (\bibinfo {year} {2017})}\BibitemShut {NoStop}%
\bibitem [{\citenamefont {Bogdanov}\ \emph {et~al.}(2019)\citenamefont
  {Bogdanov}, \citenamefont {Kazinski},\ and\ \citenamefont
  {Lazarenko}}]{Bogdanov:2019ocq}%
  \BibitemOpen
  \bibfield  {author} {\bibinfo {author} {\bibfnamefont {O.~V.}\ \bibnamefont
  {Bogdanov}}, \bibinfo {author} {\bibfnamefont {P.~O.}\ \bibnamefont
  {Kazinski}},\ and\ \bibinfo {author} {\bibfnamefont {G.~Y.}\ \bibnamefont
  {Lazarenko}},\ }\bibfield  {title} {\bibinfo {title} {Semiclassical
  probability of radiation of twisted photons in the ultrarelativistic limit},\
  }\href {https://doi.org/10.1103/PhysRevD.99.116016} {\bibfield  {journal}
  {\bibinfo  {journal} {Phys. Rev. D}\ }\textbf {\bibinfo {volume} {99}},\
  \bibinfo {pages} {116016} (\bibinfo {year} {2019})}\BibitemShut {NoStop}%
\bibitem [{\citenamefont {Chen}\ \emph {et~al.}(2019)\citenamefont {Chen},
  \citenamefont {Hatsagortsyan},\ and\ \citenamefont
  {Keitel}}]{chen2019generation}%
  \BibitemOpen
  \bibfield  {author} {\bibinfo {author} {\bibfnamefont {Y.-Y.}\ \bibnamefont
  {Chen}}, \bibinfo {author} {\bibfnamefont {K.~Z.}\ \bibnamefont
  {Hatsagortsyan}},\ and\ \bibinfo {author} {\bibfnamefont {C.~H.}\
  \bibnamefont {Keitel}},\ }\bibfield  {title} {\bibinfo {title} {Generation of
  twisted $\ensuremath{\gamma}$-ray radiation by nonlinear thomson scattering
  of twisted light},\ }\href {https://doi.org/10.1063/1.5086347} {\bibfield
  {journal} {\bibinfo  {journal} {Matter Radiat. Extremes}\ }\textbf {\bibinfo
  {volume} {4}},\ \bibinfo {pages} {024401} (\bibinfo {year}
  {2019})}\BibitemShut {NoStop}%
\bibitem [{\citenamefont {Maruyama}\ \emph {et~al.}(2025)\citenamefont
  {Maruyama}, \citenamefont {Hayakawa}, \citenamefont {Hajima}, \citenamefont
  {Kajino},\ and\ \citenamefont {Cheoun}}]{maruyama2025photon}%
  \BibitemOpen
  \bibfield  {author} {\bibinfo {author} {\bibfnamefont {T.}~\bibnamefont
  {Maruyama}}, \bibinfo {author} {\bibfnamefont {T.}~\bibnamefont {Hayakawa}},
  \bibinfo {author} {\bibfnamefont {R.}~\bibnamefont {Hajima}}, \bibinfo
  {author} {\bibfnamefont {T.}~\bibnamefont {Kajino}},\ and\ \bibinfo {author}
  {\bibfnamefont {M.-K.}\ \bibnamefont {Cheoun}},\ }\bibfield  {title}
  {\bibinfo {title} {Photon vortex generation from nonlinear compton scattering
  in feynman approach},\ }\href {https://doi.org/10.1103/PhysRevD.111.016016}
  {\bibfield  {journal} {\bibinfo  {journal} {Phys. Rev. D}\ }\textbf {\bibinfo
  {volume} {111}},\ \bibinfo {pages} {016016} (\bibinfo {year}
  {2025})}\BibitemShut {NoStop}%
\bibitem [{\citenamefont {Liao}\ \emph {et~al.}(2025)\citenamefont {Liao},
  \citenamefont {Wang},\ and\ \citenamefont {Wu}}]{liao2025all}%
  \BibitemOpen
  \bibfield  {author} {\bibinfo {author} {\bibfnamefont {Y.}~\bibnamefont
  {Liao}}, \bibinfo {author} {\bibfnamefont {Q.-Y.}\ \bibnamefont {Wang}},\
  and\ \bibinfo {author} {\bibfnamefont {Y.}~\bibnamefont {Wu}},\ }\bibfield
  {title} {\bibinfo {title} {All-vortex nonlinear compton scattering in a
  polarized laser field},\ }\href {https://doi.org/10.1103/gxs8-vgbj}
  {\bibfield  {journal} {\bibinfo  {journal} {Phys. Rev. D}\ }\textbf {\bibinfo
  {volume} {112}},\ \bibinfo {pages} {033004} (\bibinfo {year}
  {2025})}\BibitemShut {NoStop}%
\bibitem [{\citenamefont {Narozhny}\ and\ \citenamefont
  {Fofanov}(2000)}]{narozhny2000quantum}%
  \BibitemOpen
  \bibfield  {author} {\bibinfo {author} {\bibfnamefont {N.~B.}\ \bibnamefont
  {Narozhny}}\ and\ \bibinfo {author} {\bibfnamefont {M.~S.}\ \bibnamefont
  {Fofanov}},\ }\bibfield  {title} {\bibinfo {title} {Quantum processes in a
  two-mode laser field},\ }\href {https://doi.org/10.1134/1.559121} {\bibfield
  {journal} {\bibinfo  {journal} {J. Exp. Theor. Phys.}\ }\textbf {\bibinfo
  {volume} {90}},\ \bibinfo {pages} {415} (\bibinfo {year} {2000})}\BibitemShut
  {NoStop}%
\bibitem [{\citenamefont {Wistisen}(2014)}]{PhysRevD.90.125008}%
  \BibitemOpen
  \bibfield  {author} {\bibinfo {author} {\bibfnamefont {T.~N.}\ \bibnamefont
  {Wistisen}},\ }\bibfield  {title} {\bibinfo {title} {Interference effect in
  nonlinear compton scattering},\ }\href
  {https://doi.org/10.1103/PhysRevD.90.125008} {\bibfield  {journal} {\bibinfo
  {journal} {Phys. Rev. D}\ }\textbf {\bibinfo {volume} {90}},\ \bibinfo
  {pages} {125008} (\bibinfo {year} {2014})}\BibitemShut {NoStop}%
\bibitem [{\citenamefont {Taira}\ and\ \citenamefont
  {Katoh}(2018)}]{PhysRevA.98.052130}%
  \BibitemOpen
  \bibfield  {author} {\bibinfo {author} {\bibfnamefont {Y.}~\bibnamefont
  {Taira}}\ and\ \bibinfo {author} {\bibfnamefont {M.}~\bibnamefont {Katoh}},\
  }\bibfield  {title} {\bibinfo {title} {Gamma-ray vortices emitted from
  nonlinear inverse thomson scattering of a two-wavelength laser beam},\ }\href
  {https://doi.org/10.1103/PhysRevA.98.052130} {\bibfield  {journal} {\bibinfo
  {journal} {Phys. Rev. A}\ }\textbf {\bibinfo {volume} {98}},\ \bibinfo
  {pages} {052130} (\bibinfo {year} {2018})}\BibitemShut {NoStop}%
\bibitem [{\citenamefont {Bogdanov}\ \emph
  {et~al.}(2026{\natexlab{a}})\citenamefont {Bogdanov}, \citenamefont {Bragin},
  \citenamefont {Kazinski},\ and\ \citenamefont {Ryakin}}]{Bogdanov:2025vdl}%
  \BibitemOpen
  \bibfield  {author} {\bibinfo {author} {\bibfnamefont {O.}~\bibnamefont
  {Bogdanov}}, \bibinfo {author} {\bibfnamefont {S.}~\bibnamefont {Bragin}},
  \bibinfo {author} {\bibfnamefont {P.}~\bibnamefont {Kazinski}},\ and\
  \bibinfo {author} {\bibfnamefont {V.}~\bibnamefont {Ryakin}},\ }\bibfield
  {title} {\bibinfo {title} {Radiation of twisted photons in elliptical
  multifrequency undulators},\ }\href
  {https://doi.org/10.1016/j.radphyschem.2026.113701} {\bibfield  {journal}
  {\bibinfo  {journal} {Radiat. Phys. Chem.}\ }\textbf {\bibinfo {volume}
  {243}},\ \bibinfo {pages} {113701} (\bibinfo {year}
  {2026}{\natexlab{a}})}\BibitemShut {NoStop}%
\bibitem [{\citenamefont {Bogdanov}\ \emph
  {et~al.}(2026{\natexlab{b}})\citenamefont {Bogdanov}, \citenamefont {Bragin},
  \citenamefont {Kazinski},\ and\ \citenamefont {Ryakin}}]{Bogdanov:2026tbl}%
  \BibitemOpen
  \bibfield  {author} {\bibinfo {author} {\bibfnamefont {O.~V.}\ \bibnamefont
  {Bogdanov}}, \bibinfo {author} {\bibfnamefont {S.~V.}\ \bibnamefont
  {Bragin}}, \bibinfo {author} {\bibfnamefont {P.~O.}\ \bibnamefont
  {Kazinski}},\ and\ \bibinfo {author} {\bibfnamefont {V.~A.}\ \bibnamefont
  {Ryakin}},\ }\bibfield  {title} {\bibinfo {title} {{Three-frequency helical
  undulator as a source of photons in composite twisted states}},\ }\href@noop
  {} {\  (\bibinfo {year} {2026}{\natexlab{b}})},\ \Eprint
  {https://arxiv.org/abs/2602.07942} {arXiv:2602.07942} \BibitemShut {NoStop}%
\bibitem [{\citenamefont {Ritus}(1985)}]{Ritus:1985vta}%
  \BibitemOpen
  \bibfield  {author} {\bibinfo {author} {\bibfnamefont {V.~I.}\ \bibnamefont
  {Ritus}},\ }\bibfield  {title} {\bibinfo {title} {Quantum effects of the
  interaction of elementary particles with an intense electromagnetic field},\
  }\href {https://doi.org/10.1007/BF01120220} {\bibfield  {journal} {\bibinfo
  {journal} {J. Sov. Laser Res.}\ }\textbf {\bibinfo {volume} {6}},\ \bibinfo
  {pages} {497} (\bibinfo {year} {1985})}\BibitemShut {NoStop}%
\bibitem [{\citenamefont {Di~Piazza}\ \emph {et~al.}(2012)\citenamefont
  {Di~Piazza}, \citenamefont {M\"uller}, \citenamefont {Hatsagortsyan},\ and\
  \citenamefont {Keitel}}]{DiPiazza2012}%
  \BibitemOpen
  \bibfield  {author} {\bibinfo {author} {\bibfnamefont {A.}~\bibnamefont
  {Di~Piazza}}, \bibinfo {author} {\bibfnamefont {C.}~\bibnamefont {M\"uller}},
  \bibinfo {author} {\bibfnamefont {K.~Z.}\ \bibnamefont {Hatsagortsyan}},\
  and\ \bibinfo {author} {\bibfnamefont {C.~H.}\ \bibnamefont {Keitel}},\
  }\bibfield  {title} {\bibinfo {title} {Extremely high-intensity laser
  interactions with fundamental quantum systems},\ }\href
  {https://doi.org/10.1103/RevModPhys.84.1177} {\bibfield  {journal} {\bibinfo
  {journal} {Rev. Mod. Phys.}\ }\textbf {\bibinfo {volume} {84}},\ \bibinfo
  {pages} {1177} (\bibinfo {year} {2012})}\BibitemShut {NoStop}%
\bibitem [{\citenamefont {Fedotov}\ \emph {et~al.}(2023)\citenamefont
  {Fedotov}, \citenamefont {Ilderton}, \citenamefont {Karbstein}, \citenamefont
  {King}, \citenamefont {Seipt}, \citenamefont {Taya},\ and\ \citenamefont
  {Torgrimsson}}]{Fedotov2023}%
  \BibitemOpen
  \bibfield  {author} {\bibinfo {author} {\bibfnamefont {A.}~\bibnamefont
  {Fedotov}}, \bibinfo {author} {\bibfnamefont {A.}~\bibnamefont {Ilderton}},
  \bibinfo {author} {\bibfnamefont {F.}~\bibnamefont {Karbstein}}, \bibinfo
  {author} {\bibfnamefont {B.}~\bibnamefont {King}}, \bibinfo {author}
  {\bibfnamefont {D.}~\bibnamefont {Seipt}}, \bibinfo {author} {\bibfnamefont
  {H.}~\bibnamefont {Taya}},\ and\ \bibinfo {author} {\bibfnamefont
  {G.}~\bibnamefont {Torgrimsson}},\ }\bibfield  {title} {\bibinfo {title}
  {Advances in qed with intense background fields},\ }\href
  {https://doi.org/10.1016/j.physrep.2023.01.003} {\bibfield  {journal}
  {\bibinfo  {journal} {Phys. Rep.}\ }\textbf {\bibinfo {volume} {1010}},\
  \bibinfo {pages} {1} (\bibinfo {year} {2023})}\BibitemShut {NoStop}%
\bibitem [{sup()}]{supplemental}%
  \BibitemOpen
  \href@noop {} {}\bibinfo {note} {See the Supplemental Material for the
  derived expressions of the $S$-matrix element and the vortex photon
  wavefunction, as well as numerical results for different laser-field
  configurations and photon helicities.}\BibitemShut {Stop}%
\bibitem [{\citenamefont {Jentschura}\ and\ \citenamefont
  {Serbo}(2011{\natexlab{b}})}]{jentschura2011compton}%
  \BibitemOpen
  \bibfield  {author} {\bibinfo {author} {\bibfnamefont {U.~D.}\ \bibnamefont
  {Jentschura}}\ and\ \bibinfo {author} {\bibfnamefont {V.~G.}\ \bibnamefont
  {Serbo}},\ }\bibfield  {title} {\bibinfo {title} {Compton upconversion of
  twisted photons: backscattering of particles with non-planar wave
  functions},\ }\href {https://doi.org/10.1140/epjc/s10052-011-1571-z}
  {\bibfield  {journal} {\bibinfo  {journal} {Eur. Phys. J. C}\ }\textbf
  {\bibinfo {volume} {71}},\ \bibinfo {pages} {1571} (\bibinfo {year}
  {2011}{\natexlab{b}})}\BibitemShut {NoStop}%
\bibitem [{\citenamefont {Bliokh}\ \emph {et~al.}(2017)\citenamefont {Bliokh},
  \citenamefont {Ivanov}, \citenamefont {Guzzinati}, \citenamefont {Clark},
  \citenamefont {Van~Boxem}, \citenamefont {B{\'e}ch{\'e}}, \citenamefont
  {Juchtmans}, \citenamefont {Alonso}, \citenamefont {Schattschneider},
  \citenamefont {Nori} \emph {et~al.}}]{bliokh2017theory}%
  \BibitemOpen
  \bibfield  {author} {\bibinfo {author} {\bibfnamefont {K.}~\bibnamefont
  {Bliokh}}, \bibinfo {author} {\bibfnamefont {I.}~\bibnamefont {Ivanov}},
  \bibinfo {author} {\bibfnamefont {G.}~\bibnamefont {Guzzinati}}, \bibinfo
  {author} {\bibfnamefont {L.}~\bibnamefont {Clark}}, \bibinfo {author}
  {\bibfnamefont {R.}~\bibnamefont {Van~Boxem}}, \bibinfo {author}
  {\bibfnamefont {A.}~\bibnamefont {B{\'e}ch{\'e}}}, \bibinfo {author}
  {\bibfnamefont {R.}~\bibnamefont {Juchtmans}}, \bibinfo {author}
  {\bibfnamefont {M.}~\bibnamefont {Alonso}}, \bibinfo {author} {\bibfnamefont
  {P.}~\bibnamefont {Schattschneider}}, \bibinfo {author} {\bibfnamefont
  {F.}~\bibnamefont {Nori}}, \emph {et~al.},\ }\bibfield  {title} {\bibinfo
  {title} {Theory and applications of free-electron vortex states},\ }\href
  {https://doi.org/10.1016/j.physrep.2017.05.006} {\bibfield  {journal}
  {\bibinfo  {journal} {Phys. Rep.}\ }\textbf {\bibinfo {volume} {690}},\
  \bibinfo {pages} {1} (\bibinfo {year} {2017})}\BibitemShut {NoStop}%
\bibitem [{\citenamefont {Wong}\ \emph {et~al.}(2021)\citenamefont {Wong},
  \citenamefont {Rivera}, \citenamefont {Murdia}, \citenamefont {Christensen},
  \citenamefont {Joannopoulos}, \citenamefont {Soljačić},\ and\ \citenamefont
  {Kaminer}}]{Wong2021}%
  \BibitemOpen
  \bibfield  {author} {\bibinfo {author} {\bibfnamefont {L.~J.}\ \bibnamefont
  {Wong}}, \bibinfo {author} {\bibfnamefont {N.}~\bibnamefont {Rivera}},
  \bibinfo {author} {\bibfnamefont {C.}~\bibnamefont {Murdia}}, \bibinfo
  {author} {\bibfnamefont {T.}~\bibnamefont {Christensen}}, \bibinfo {author}
  {\bibfnamefont {J.~D.}\ \bibnamefont {Joannopoulos}}, \bibinfo {author}
  {\bibfnamefont {M.}~\bibnamefont {Soljačić}},\ and\ \bibinfo {author}
  {\bibfnamefont {I.}~\bibnamefont {Kaminer}},\ }\bibfield  {title} {\bibinfo
  {title} {Control of quantum electrodynamical processes by shaping electron
  wavepackets},\ }\href {https://doi.org/10.1038/s41467-021-21367-1} {\bibfield
   {journal} {\bibinfo  {journal} {Nat. Photonics}\ }\textbf {\bibinfo {volume}
  {12}},\ \bibinfo {pages} {1700} (\bibinfo {year} {2021})}\BibitemShut
  {NoStop}%
\end{thebibliography}%
\end{document}